%% file: main.tex
\documentclass[%
amsmath,aps,showkeys,
amssymb,twocolumn,pra,
superscriptaddress,
sort&compress,
]{revtex4-2}

\usepackage{url}
\usepackage{hyperref}
\hypersetup{
    colorlinks,
    linkcolor={black},
    citecolor={blue!80!black},
    urlcolor={blue!80!black},
    breaklinks=true
}
\newcommand{\apscopyright}{© 2023 American Physical Society}
\newcommand{\apscitation}{Phys. Rev. A 108, 053110 (2023), \href{https://doi.org/10.1103/PhysRevA.108.053110}{DOI: 10.1103/PhysRevA.108.053110}}

\usepackage{graphicx}% Include figure files
\usepackage{bm}% bold math
\usepackage{bbold}
\usepackage{dcolumn}% Align table columns on decimal point
\newcolumntype{d}[1]{D{.}{.}{#1}}

\usepackage[dvipsnames]{xcolor}

\newcommand{\abs}[1]{\left\lvert{#1}\right\rvert}
\newcommand{\norm}[1]{{\left\lVert{#1}\right\rVert}}

\renewcommand{\b}[1]{\langle#1|} % bra
\renewcommand{\k}[1]{|#1\rangle} % ket
\newcommand{\com}[2]{\left[#1,#2\right]}
\newcommand{\proj}[1]{|#1\rangle\langle #1|} % projector |#1><#1| onto #1
\newcommand{\rmi}{\mathrm{i}}

\newcommand{\R}{\rangle}
\renewcommand{\L}{\langle}
\renewcommand{\d}{\mathrm{d}}
\def\tr{\mbox{tr}}

\usepackage{url}
\usepackage{tikz}
\usepackage{pgfplots}
\pgfplotsset{compat=1.14}
\usepackage{booktabs}
\usepackage{pgfplotstable}
\usepackage{float}
\usepackage{subcaption}
\renewcommand{\arraystretch}{1.4} 
\usepackage{multirow}
\usepackage{bbold}
\usepackage{listings}

\usepackage{enumerate}
\usepackage{latexsym}
\usepackage{amsfonts, amsmath, bbm, amssymb, amsthm}
\usepackage{epic}

\makeatletter
\g@addto@macro{\UrlBreaks}{\do\/\do\-}
\makeatother

\begin{document}

\title{Nuclear electric resonance for spatially-resolved spin control \\ via pulsed optical excitation in the UV-visible spectrum}
\thanks{\apscopyright. The published version is available at \apscitation.}

\author{Johannes K. Krondorfer}
\affiliation{ 
Institute of Experimental Physics, Graz University of Technology, Petersgasse 16, 8010 Graz
}

\author{Andreas W. Hauser}%
\email{andreas.w.hauser@gmail.com}
\affiliation{ 
Institute of Experimental Physics, Graz University of Technology, Petersgasse 16, 8010 Graz
}

\date{\today}

\begin{abstract}
Nuclear electric resonance (NER) spectroscopy is currently experiencing a revival as a tool for nuclear spin-based quantum computing. Compared to magnetic or electric fields, local electron density fluctuations caused by changes in the atomic environment provide a much higher spatial resolution for the addressing of nuclear spins in qubit registers or within a single molecule. In this article, we investigate the possibility of coherent spin control in atoms or molecules via nuclear quadrupole resonance from first principles. An abstract, time-dependent description is provided which entails and reflects on commonly applied approximations. This formalism is then used to propose a new method we refer to as `optical' nuclear electric resonance (ONER). It employs pulsed optical excitations in the UV-visible light spectrum to modulate the electric field gradient at the position of a specific nucleus of interest by periodic changes of the surrounding electron density. Possible realizations and limitations of ONER for atomically resolved spin manipulation are discussed and tested on $^9$Be as an atomic benchmark system via electronic structure theory.
\end{abstract}


\maketitle

\section{Introduction}
Quantum technologies are attracting increasing attention in recent years, above all the field of quantum computing. This interest stems from the potential of quantum systems to outperform classical computers in tasks such as the simulation of complex quantum systems, optimization problems, or cryptography \cite{quantum_super0,quantum_super1,quantum_super2,quantum_super3,quantum_super4}. Current paradigms of quantum computers are superconducting circuits \cite{quantum_superconducting1,quantum_superconducting2}, trapped ions \cite{trapped_ions1,trapped_ions2} or atoms \cite{neutral_atoms1,neutral_atoms2}, solid-state systems such as semiconductors~\cite{semi1,semi2} or topological qubits~\cite{topo1,topo2} and nuclear spin-based quantum systems~\cite{quantum_nuclear, quantum_nuclear2,original_nature}.

A critical factor for the realization of quantum computers is the coherence time of a single qubit. Among the many systems being studied, nuclear spins are of particular interest due to their comparably large coherence time. In large ensembles, their control and detection via magnetic resonance is widely exploited. Early proposals for solid-state quantum computers utilized nuclear magnetic resonance to realize quantum search and factoring algorithms~\cite{nmr_factor,nmr_search,nmr_factor2}. However, despite the success, possible applications are intrinsically limited by the fact that oscillating magnetic fields cannot be easily confined or screened at the nanoscale. As a consequence, identical nuclear spins within a large region respond to the same signal and cannot be addressed individually. This presents a challenge for the up-scaling and the integration of nuclear systems into multi-spin devices based on magnetic control only.

Control via electric fields, on the other hand, would resolve this problem, since electric fields can be efficiently routed and confined via industrial standard procedures. Recently, there has been significant progress in using the electron-nuclear hyperfine interaction to transduce electric signals into magnetic fields for nuclear spin control~\cite{neutral_atoms2,hyperfine}, and first universal gates have been realized in this way for trapped ytterbium atoms~\cite{neutral_atoms2}. Yet, although working in principle, this type of coupling also opens a channel for nuclear spin decoherence. To maintain maximum coherence and allow for individual control of the nuclear spins, a direct, exclusively electrical control over spin states might turn out as a viable solution. Here, the use of radio frequency electric fields is believed to be suitable for an up-scaling of nuclear spin-based quantum devices, taking advantage of the nuclear quadrupole interaction (NQI), a well-known effect leading to line shifts of nuclear magnetic resonance (NMR) signals. Despite much earlier suggestions of coherent quadrupole coupling~\cite{bloembergen1961linear}, a first experimental demonstration of a controlled spin manipulation succeeded only very recently: Avoiding the mediation via a magnetic field, a coherence time of 0.1 s could be achieved for a high-spin nucleus in silicon~\cite{original_nature}. Motivated by these findings, we provide a comprehensive theoretical analysis and discuss commonly applied phenomenological approximations that have been used to either describe nuclear electric resonance (NER) or nuclear acoustic resonance (NAR)~\cite{original_nature, NAR1, NAR2}. Based on this theoretical framework, we then propose a new protocol of nuclear spin control using electric fields in the visible regime, a technique we refer to as optical nuclear electric resonance (ONER). Bringing optical excitation into play, the entire field of optoelectronics and nanophotonics might enter the quest for viable quantum computing technologies based on nuclear spin processes.

Our article is structured as follows. In a first step, we introduce a nuclear quadrupole Hamiltonian for a molecular system and describe the interaction of the nuclear quadrupole moment with the electric field gradient (EFG). Relevant properties, such as energy correction terms and transition elements of the quadrupole Hamiltonian, are investigated for an open two-level system. Second, we derive an abstract description of time-dependent nuclear quadrupole interaction for NER and NAR from quantum mechanical principles. In a third step, we use our formalism to propose a protocol for ONER as a new paradigm and apply it to a single beryllium atom in a benchmark study.

\section{Methods}
\label{sec:methods}

\subsection{Nuclear quadrupole Hamiltonian} \label{sec:nuc_quad_ham}
Atomic nuclei consist of protons and neutrons and therefore exhibit a charge distribution. The latter has a vanishing dipole moment with respect to the center of charge of the nucleus, but might feature a non-vanishing quadrupole moment, which is related to the nuclear spin $I$ and interacts with the electric field gradient (EFG), the second derivative of the electric potential, at the position of the nucleus. The corresponding Hamiltonian can be written as
\begin{align} \label{eq:phi2q}
\begin{split}
H_Q = I_\mu Q_{\mu\nu} I_\nu,
\end{split}
\end{align}
with $Q_{\mu\nu} = \frac{q}{2I(2I-1)} \Phi_{\mu\nu}$, where $\Phi_{\mu\nu}$ is the EFG tensor and $q$ is the scalar quadrupole moment of the nucleus. Note that we implicitly sum over double appearing Greek indices in this expression, a convention we keep throughout the manuscript. A detailed derivation of this Hamiltonian can be found in Appendix~\ref{sec:A1}. The numerical values of the scalar quadrupole moments of different nuclei are tabulated in Refs.~\cite{stone2005table,tabel_of_quadrupole_moments}. Note that only nuclei with nuclear spin $I > \frac{1}{2}$ can have a non-vanishing quadrupole tensor, as indicated by the proportionality constant. We will refer to the tensor $Q$ as NQI tensor in the further discussion.

\subsubsection{Quadrupole energy splitting}\label{sec:quad_ham_perturb}
Before continuing with the derivation of a suitable model, it is convenient to investigate a few properties of the spin quadrupole Hamiltonian. Although quadrupole splitting also occurs if no external magnetic field is present, it is reasonable to apply an external magnetic field to the nucleus of interest in order to obtain a sufficiently large splitting of the nuclear spin states. In an external magnetic field $B = B_0 e_z$, the spin Hamiltonian for a quadrupolar nucleus reads
\begin{align} \label{eq:HbHq}
H = H_B + H_Q = -\gamma_n B_0 I_z + I_\mu Q_{\mu\nu} I_\nu,
\end{align}
with $\gamma_n$ denoting the gyromagnetic moment of the nucleus. If the energy splitting due to the magnetic field is large compared to the quadrupole line splitting, the energy correction due to NQI can be treated perturbatively. In good approximation, the eigenstates of the total Hamiltonian can be described by the eigenstates of the Zeeman-Hamiltonian, which are just the orientational nuclear spin states $\k{m_I}$ for a fixed spin quantum number $I$. In first order, the corrected energies are given by
\begin{align}
\begin{split}
    \mathcal{E}_{m_I}^{(1)} 
    &= \b{m_I} H \k{m_I} \\
    &= -\gamma_n B_0 m_I + \left(\frac{3m_I^2}{2} - \frac{I(I+1)}{2} \right) Q_{zz}.
\end{split}
\end{align}
This leads to corrected transition energies of
\begin{align} \label{eq:trans_energy_corr}
\begin{split}
\Delta \mathcal{E}(m_I-1 \rightarrow m_I) &= -\gamma_n B_0 + \frac{3}{2}(2m_I - 1)Q_{zz} \\
\Delta \mathcal{E}(m_I-2 \rightarrow m_I) &= -2\gamma_n B_0 + \frac{3}{2}(4m_I - 4)Q_{zz}. 
\end{split}
\end{align}
Note that this correction opens the possibility to address specific transitions individually, which is not possible in the case of equidistant Zeeman-splitting alone. These corrected transition energies become relevant when choosing a suitable driving frequency for the nuclear spin system.

\subsubsection{Transition elements and time-dependent couplings}\label{sec:trans_el}
In order to drive transitions between different spin states $\k{m_I}$ via direct quadrupole coupling a time-dependent variation of the EFG tensor is necessary. In the following, we calculate the transition elements for a quadrupolar nucleus in an external, constant magnetic field $B = B_0 e_z$, similar to the situation discussed in the supplementary material of  Ref.~\cite{original_nature}. The spin Hamiltonian for a nucleus of spin $I$, subject to a time-dependent NQI tensor $Q_{\mu\nu}(t)$, is given by~\eqref{eq:HbHq}. The time-dependence is inherited from the EFG tensor, which can be manipulated. The Rabi frequency for transitions from some $m_I$ to $m_I^{\prime}$ is mainly determined by the transition amplitudes, denoted as $g_{m_I \rightarrow m_I^{\prime}}$ below.

Any quadratic combination of ${x,y,z}$ components of the spin operators can appear in the interaction Hamiltonian, but it is immediately clear that only transitions $m_I \rightarrow m_I \pm 1$ and $m_I \rightarrow m_I \pm 2$ are possible since the interaction is quadratic in the spin operators. Transitions with $\Delta m_I = \pm 1$ are driven by the terms $I_x I_z$, $I_y I_z$ and their corresponding adjoints. Since the quadrupole interaction is symmetric, we can combine the terms to derive the proportionality factor of $Q_{xz}(t)$. For $m_I \rightarrow m_I -1 $ transitions one obtains the transition amplitude
\begin{align} \label{eq:trans1}
\begin{split}
g_{m_I \rightarrow m_I-1}(t) &= \alpha_{m_I-1 \leftrightarrow m_I} \left( Q_{xz}(t) + \rmi Q_{yz}(t) \right);\\
\alpha_{m_I-1 \leftrightarrow m_I} &= \frac{1}{2} \abs{2 m_I -1} \sqrt{I(I+1) - m_I (m_I - 1)}.
\end{split}
\end{align}
Transitions with $\Delta m_I = \pm 2$ are driven by the terms quadratic in the $x$ and $y$ spin operators, i.e. $I_x^2$, $I_y^2$, $I_xI_y$ and $I_yI_x$. Calculating the transition amplitude for the corresponding $m_I \rightarrow m_I -2 $ transitions yields
\begin{align} \label{eq:trans2}
\begin{split}
g_{m_I \rightarrow m_I-2}(t) &=\! \beta_{m_I-2 \leftrightarrow m_I} \left( Q_{xx}(t) - Q_{yy}(t) +  2\rmi Q_{yx}(t) \right) \\
\beta_{m_I-2 \leftrightarrow m_I} &= \frac{1}{4} \sqrt{ \left( I(I+1) - (m_I-1) ( m_I -2) \right) } \\ &\qquad\quad \times \sqrt{ \left( I(I+1) - m_I ( m_I -1) \right)}.
\end{split}
\end{align}
Note that the modulus of these transition amplitudes will determine the Rabi frequency of the respective spin level transitions in the dynamics simulations later.

\subsection{The open two-level system}\label{sec:2L_sys}
In order to derive a protocol for ONER we consider pulsed excitations of a two-level system to drive the nuclear transitions. The energy of the ground state $\k{g}$ is set to zero, the energy of the excited state $\k{e}$ is denoted as $\omega_0$. We model the interaction in the presence of an external electric field $E(t) = E_0\cos(\omega t) \hat{\varepsilon}$ of amplitude $E_0$ and linear polarization $\hat{\varepsilon}$. The detuning of the electric field is denoted as $\Delta = \omega - \omega_0$. Within the dipole approximation, the dynamics is described by the Hamiltonian
\begin{equation} \label{eq:2level_ham}
    H_{2L} = \omega_0 \sigma^\dagger\sigma - \mathcal{P}\cdot E(t),
\end{equation}
with  $\mathcal{P} = \left(\mu\sigma + \mu^*\sigma^\dagger\right)$ as dipole transition operator, with $\sigma = \k{g}\b{e}$ as the lowering operator and $\mu$ as the corresponding dipole transition matrix element. It is standard practice to perform the rotating wave approximation, neglecting fast oscillating terms in the Hamiltonian, and to transform the system in the rotating frame by a unitary transformation~\cite{quantum_optics_long_book}. This yields a time-independent Hamiltonian
\begin{equation}\label{eq:2level_ham_rot}
    H' = -\Delta\sigma^\dagger\sigma + \frac{\Omega}{2}\sigma + \frac{\Omega^*}{2}\sigma^\dagger,
\end{equation}
with $\Omega = - \b{g}\mathcal{P} \cdot \hat{\varepsilon}\k{e}E_0$ denoting the Rabi frequency, which depends on the relative orientation of the dipole moment and the polarization of the electric field.

If interactions with an environment are taken into account, decay and dephasing terms may enter through a Linblad Master equation (a brief introduction to open quantum systems and density operators can be found in Appendix~\ref{sec:A2}). Switching to the density operator formalism, the general form reads
\begin{equation}\label{eq:2L_master_eq}
    \rmi\hbar\partial_t\rho' = \com{H'}{\rho'} + \rmi\hbar\Gamma\mathcal{L}[\sigma]\rho' + \rmi\hbar\frac{\gamma_c}{2}\mathcal{L}[\sigma_z]\rho', 
\end{equation}
with $\sigma$ inducing decays with rate $\Gamma$ from the excited state to the ground state and $\sigma_z = \proj{e} - \proj{g}$ inducing coherence loss, quantified by the decay rate $\gamma_c$. The superoperator $\mathcal{L}$ is defined as
\begin{equation} \label{eq:super_op}
    \mathcal{L}[c]\rho_S = c\rho_S c^\dagger - \frac{1}{2}\left( c^\dagger c \rho_S + \rho_S c^\dagger c \right),
\end{equation}
for some collapse operator $c$.

The representation as Master equation is convenient for a general numerical implementation and generalizations to more complicated systems. Possible solutions can be obtained numerically, e.g. via the Python library QuTiP~\cite{qutip1,qutip2}.

Whereas the isolated system does not have any steady-state solutions, the open system tends toward an equilibrium for $t\rightarrow\infty$. Demanding $\partial_t \rho = 0$ and solving the remaining homogeneous linear equation, one obtains the steady-state solutions for the density operator matrix elements,
\begin{align}\label{eq:2level_oo}
    \begin{split}
        \rho_{ee}(t\rightarrow\infty) &= \frac{\Omega^2}{2\gamma_\perp\Gamma}\; \frac{1}{1+\frac{\Delta^2}{\gamma_\perp^2} + \frac{\Omega^2}{\gamma_\perp\Gamma}}\\
        \rho'_{eg}(t\rightarrow\infty) &= \frac{\rmi\Omega}{2\gamma_\perp}\; \frac{1 + \frac{\rmi\Delta}{\gamma_\perp}}{1+\frac{\Delta^2}{\gamma_\perp^2} + \frac{\Omega^2}{\gamma_\perp\Gamma}}, 
    \end{split}
\end{align}
with the definition $\gamma_\perp = \frac{\Gamma}{2} + \gamma_c$.
These solutions, and thus the dynamics of the system, depend on the ratio of Rabi frequency and the decay rate. The higher the decay rate, the faster the convergence to the steady-state solution.

\section{Results and Discussion}
\label{sec:results}

\subsection{Nuclear electric resonance}
Having established the interaction principles of a quadrupolar nucleus and the EFG, we are now interested in the possibility of controlling the nuclear spin coherently. This will be achieved through a modulation of the EFG at the position of the nucleus we aim to address. We start by a general formulation of the quantum mechanical equations and derive an evolution equation for the spin system. For the sake of a reduced formalism, only a single nucleus will be assumed; a generalization of this interaction to systems containing several nuclei is straight-forward.

\subsubsection{General description} \label{sec:NER_general}
We consider a generic molecular system exposed either to external strain in case of NAR or to external fields in the case of NER. All non-spin related contributions to the Hamiltonian are collected in a `molecular' part, i.e. kinetic energies of all particles and Coulomb interactions plus external strain or external electric field effects, and a `spin' part including spin interactions with the external magnetic field and with the electric field gradient. In compact form, the total Hamiltonian reads
\begin{align}
\begin{split}
    H(t) 
    &= H_M(t) \otimes \mathbb{1} + \mathbb{1} \otimes H_B \\
    &\qquad+ Q_{\mu\nu} \otimes I_\mu I_\nu,
\end{split}
\end{align}
with $H_M(t)$ denoting the molecular Hamiltonian, $H_B$ denoting the Zeeman interaction Hamiltonian of the nuclear spin, and $Q_{\mu\nu}$ as the NQI tensor derived in Section~\ref{sec:nuc_quad_ham}. The molecular part might also contain coupling terms to the environment in form of Lindblad superoperators that only act on the molecular system. This accounts for the fact that decay rates of electronic states are typically very fast compared to time scales of the nuclear spin system. Note that the two separate Hilbert spaces are coupled only through the nuclear quadrupole interaction. For the purpose of this work, we assume that additional spin interactions, such as hyperfine coupling or spin-spin coupling with neighboring atoms, are negligible, and will choose our benchmark system accordingly. A similar analysis might be possible considering additional interactions; however, this is much more involved, and might even hinder coherent control via quadrupole interaction, as was suggested by Ref.~\cite{original_nature}. Since the coupling between molecular system and spin system is small, we will further neglect any reverse impact of the spin system onto the molecular system in the time evolution of the molecular system. This step greatly simplifies the computational treatment, since all relevant properties of the isolated molecular system become accessible via standard computational chemistry packages. In this work, we will use the Molpro suite of programs~\cite{Molpro,Molpro1,Molpro2} for the computation of EFG fluctuations caused by electronic excitation.

Since the coupled system obeys the von Neumann equation, we have to take partial traces to obtain dynamical equations for the molecular system and the spin system, respectively. For details, we refer to Appendix~\ref{sec:A2}. Supposing a weak coupling only due to quadrupole interaction, i.e. $\norm{Q}:= \max_{\mu\nu}\abs{Q_{\mu\nu}}$, we may assume that the total density operator remains decomposable over time, i.e. that the Born approximation
\begin{equation}
    \rho(t) = \rho_M(t)\otimes\rho_S(t)
\end{equation}
holds throughout time evolution, where $\rho_M$ and $\rho_S$ are the partial density operators of the molecular system and the spin system, respectively. This approximation is reasonable since the coupling strength of the spin system is negligibly small compared to the energy scale of the molecular Hamiltonian. With these approximations in place, we obtain an evolution equation for the molecular part of the form
\begin{align}
\begin{split}
    \rmi\partial_t\rho_{M} &= \rmi\partial_t\tr_S{\left\{\rho\right\}}
    = \tr_S\left\{\com{H}{\rho}\right\}\\
    &\approx \com{H_M(t)}{\rho_{M}} \overbrace{\tr_S{\left\{\rho_S\right\}}}^{=1} + \rho_{M} \overbrace{\tr_S{\left\{\com{H_B}{\rho_S}\right\}}}^{=0} \\
    &\qquad\qquad + \com{Q_{\mu\nu}}{\rho_{M}}\tr_S{\left\{I_\mu I_\nu \rho_S\right\}} \\
    &= \com{H_{M}(t)}{\rho_{M}} + \mathcal{O}\left(\norm{Q}\right).
\end{split}
\end{align}
Note that the error of the molecular density operator of the isolated system with respect to the coupled system is of the order of $\mathcal{O}\left(\norm{Q}\right)$. Furthermore, we obtain an evolution equation for the spin part by taking the partial trace over the molecular part, $\tr_M$, which yields 
\begin{align} \label{eq:dyn_molecular_spin}
\begin{split}
    \rmi\partial_t\rho_{S} &= \rmi\partial_t\tr_{M}{\left\{\rho\right\}} 
    = \tr_{M}\left\{\com{H}{\rho}\right\} \\
    &= \rho_S \overbrace{\tr_{M}{\left\{\com{H_{M}(t)}{\rho_{M}}\right\}}}^{=0} + \com{H_B}{\rho_S}\overbrace{\tr_{M}{\left\{\rho_{M}\right\}}}^{=1} \\
    &\qquad\qquad + \com{I_\mu I_\nu}{\rho_S} \underbrace{\tr_{M}{\left\{Q_{\mu\nu}\rho_{M}\right\}}}_{=\L Q_{\mu\nu} \R (t)} \\
    &= \com{H_B + \L Q_{\mu\nu} \R (t) I_\mu I_\nu}{\rho_S}.
\end{split}
\end{align}
Hence, we are left with the two dynamical equations
\begin{align} \label{eq:dyn_mol_efg}
\begin{split}
    \rmi\partial_t\rho_{M} &= \com{H_{M}(t)}{\rho_{M}} + \mathcal{O}\left(\norm{Q}\right) \\
    \rmi\partial_t\rho_{S} &= \com{H_B + \L Q_{\mu\nu} \R (t) I_\mu I_\nu}{\rho_S}, 
\end{split}
\end{align}
with
\begin{align} \label{eq:dyn_mol_efg_2}
\begin{split}
    \L Q_{\mu\nu} \R (t) &= \tr_{M}{\left\{\rho_{M}(t) Q_{\mu\nu}\right\}} + \mathcal{O}\left(\norm{Q}^2\right).
\end{split}
\end{align}
Both can be solved via pure states, i.e. via a Schrödinger equation instead of the von Neumann ansatz for density operators.

\subsubsection{Solution for the molecular system}
According to~\eqref{eq:dyn_mol_efg} and~\eqref{eq:dyn_mol_efg_2}, the molecular equation needs to be solved first, since the spin part can only be solved once the time-dependence of the NQI tensor is known. Depending on the actual problem setting the solution of the molecular system necessitates further approximations. Since the time-dependence of the external field in the molecular Hamiltonian in NER as well as NAR is slow compared to the characteristic time of electron movement, an adiabatic behavior may be assumed to obtain the time-dependence of the molecular system. A discussion of the units and orders of magnitude of NQI parameters is given in Appendix~\ref{sec:A3}. Employing the adiabatic theorem~\cite{adiabaticQM}, the wavefunction of the molecular system can be written as
\begin{equation}
    \k{\psi_M(t)} = e^{-\rmi\gamma(t)} e^{-\rmi\int_0^t \mathcal{E}(\tau) \,\d\tau}  \k{\phi_M(t)},
\end{equation}
with $\gamma(t)$ as a time-dependent phase and $\k{\phi_M(t)}$, $\mathcal{E}(t)$ as eigenfunction and eigenvalue, respectively, of the adiabatic Hamiltonian equation
\begin{equation}
    H_M(t) \k{\phi_M(t)} = \mathcal{E}(t) \k{\phi_M(t)}.
\end{equation}
Choosing the adiabatic ground state due to environmental coupling and short relaxation times in comparison to the long timescale of the nuclear quadrupole interaction, the time-dependence of the NQI tensor may be written as an expectation value of the corresponding electronic wavefunction,
\begin{equation}
    \L Q_{\mu\nu} \R(t) = \b{\psi_M(t)} Q_{\mu\nu} \k{\psi_M(t)},
\end{equation}
which can be easily calculated via common electronic structure methods, which provide the EFG tensor components $\Phi_{\mu\nu}$  (see equation~\eqref{eq:phi2q}) at the position of the nucleus.

\subsection{Optical nuclear electric resonance}\label{sec:oNER}
With all prerequisites in place, we are now in the position to propose our protocol for an optical stimulation of nuclear spin processes via pulsed light. As a special case of the above, we consider an electronic two-level system coupled to a nuclear spin system via quadrupole interaction. The environment needs to be taken into account, since typical lifetimes of electronically excited states of atoms are in the range of 10$^{-9}$~seconds, i.e. are much smaller than the typical timescales for nuclear spin control via quadrupole interaction (10$^{-3}$ to 10$^{-6}$s).

\subsubsection{General derivation of ONER}
The Hamiltonian of a two-level system coupled to a nuclear spin in an external constant magnetic field $B = B_0 e_z$ and time-dependent electric field $E(t)$ is given by
\begin{align} \label{eq:ham_2L_spin}
\begin{split}
    H &= \left(H_{2L} + H_E(t)\right)\otimes\mathbb{1} + \mathbb{1}\otimes H_B + H_Q \\
    &= \left(\omega_0\proj{e} - \mathcal{P}\cdot E(t) \right) \otimes \mathbb{1} \\
    &\qquad- \mathbb{1} \otimes \gamma_n B_0 I_z + Q_{\mu\nu} \otimes I_\mu I_\nu,
\end{split}
\end{align}
with all constants defined as above. Note that the electric field gradient, and thus the quadrupole coupling tensor $Q$, is an operator in the Hilbert space of the two-level system, i.e. a $2\times 2$ matrix in the basis $\{\k{g},\k{e}\}$ containing blocks of EFG tensors for the spin system. Only the last term in~\eqref{eq:ham_2L_spin} couples the two-level system and the nuclear spin system. The dynamics of the two-level system is mainly influenced by the time-dependent (dipole) Hamiltonian, inducing transitions of the two-level system. It will be treated via a Born-Markov Master equation with a decay constant $\Gamma \gg \abs{\gamma_n B_0},\, \norm{Q}$; see Section~\ref{sec:2L_sys} for details. This is reasonable since the decay frequency lies in the range of GHz for non-metastable excited states of isolated or weakly interacting atoms, whereas the Zeeman splitting and the quadrupole splitting are in the  MHz and kHz regime, respectively. The Rabi frequency of the two-level system is chosen to be of the order of GHz, which can be adjusted by the intensity of the laser field. The energy range of the electronic excitation lies in the order of PHz. Thus, we have established the necessary conditions for our protocol,
\begin{equation}\label{eq:2L_condition}
    \omega \approx \omega_0 \gg \Gamma \sim \Omega  \gg \abs{\gamma_n B_0},\,\norm{Q}.
\end{equation}

\begin{figure*}[ht]
    \centering
    \includegraphics[width=0.8\textwidth]{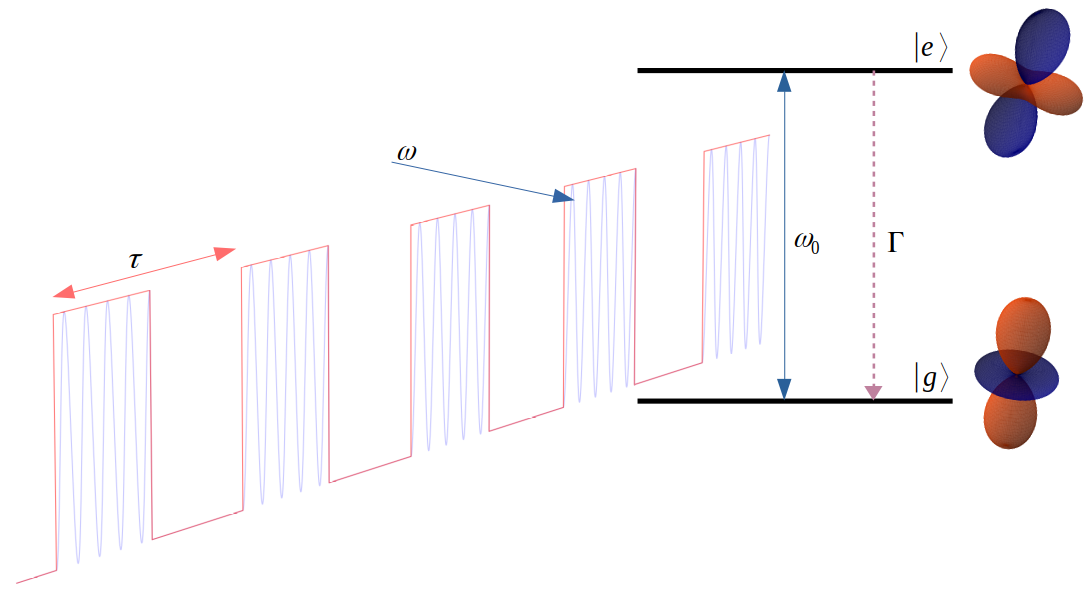}
    \caption{Schematic illustration of a pulsed excitation of a two-level system $\{\k{g},\k{e}\}$ with frequency $\omega_0$, decay rate $\Gamma$, external pulse duration $\tau$ and external field frequency $\omega$. For visibility, the frequency of the driving field is scaled. The EFG tensors of ground and excited state are illustrated at the respective state level. More information on the visualization of EFG tensors can be found in Appendix~\ref{sec:A4}.}
    \label{fig:2level_EFG_scheme}
\end{figure*}

\begin{figure*}[ht]
    \centering
    \begin{subfigure}[b]{0.49\textwidth}
        \centering
        \includegraphics[width=\textwidth]{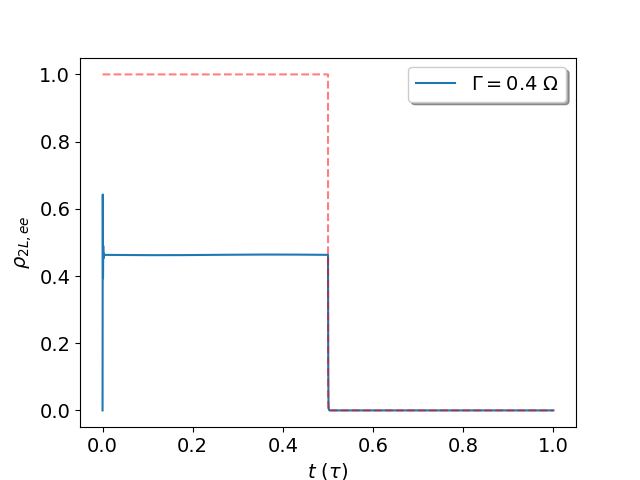}
        \caption{}
    \end{subfigure}
    \begin{subfigure}[b]{0.49\textwidth}
        \centering
        \includegraphics[width=\textwidth]{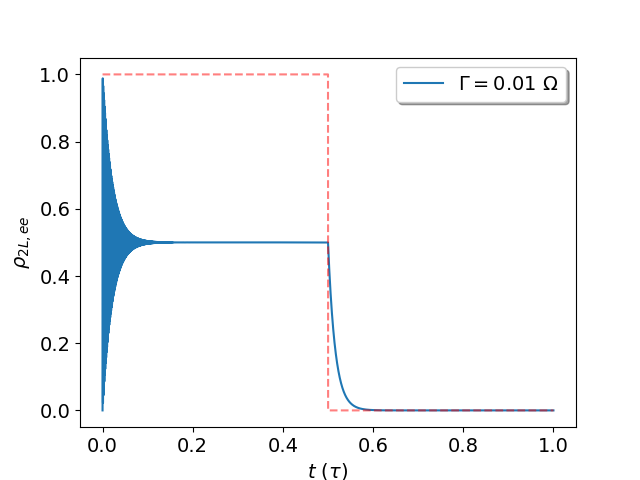}
        \caption{}
    \end{subfigure}
    \begin{subfigure}[b]{0.49\textwidth}
        \centering
        \includegraphics[width=\textwidth]{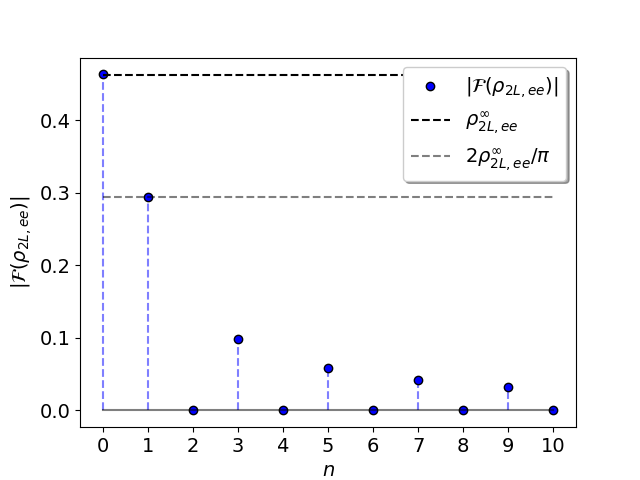}
        \caption{}
    \end{subfigure}
    \begin{subfigure}[b]{0.49\textwidth}
        \centering
        \includegraphics[width=\textwidth]{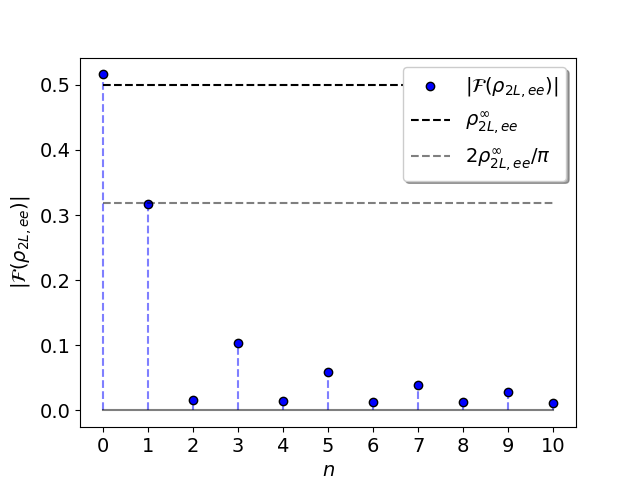}
        \caption{}
    \end{subfigure}
    \caption{Population of the excited state of a two-level system under pulsed excitation with decay with Rabi-frequency $\Omega \gg \frac{1}{\tau}$. The dashed red line in panels (a) and (b) indicates the enveloping function of the laser pulse. The $x$-axes shows one pulse duration $\tau$. Panel (a) shows the case of a large decay rate, $\Gamma \gg \frac{1}{\tau}$, whereas a moderate decay rate is shown in panel (b). In both cases the steady state solution is reached. In panel (b) the excited state population oscillates in the beginning. After turning off the external pulse, the system relaxes into the ground state, according to their respective decay rate. The Fourier analysis of the respective populations and a comparison with the steady-state solution, i.e. a perfect square signal, is shown in panels (c) and (d). The $x$-axis refers to the Fourier component $n$, with corresponding frequency $\omega_n = \frac{2\pi n}{\tau}$. It can be seen that the steady-state solution is a good approximation for the dominant Fourier modes in both cases. This becomes relevant in the theoretical analysis of the ONER protocol.}
    \label{fig:2level_EFG_pulse}
\end{figure*}

The effect of the spin system on the two-level system is negligible, as the dominant decay channel $\Gamma$ of the two-level system is governed by the interaction with the environment and the dynamics of the two-level system is much faster than the dynamics of the spin system. This also justifies the Born-like assumption that the total density matrix of the two-level system and the spin system remains decomposable throughout time-evolution. Thus, the same derivation as above is applicable, and by taking the respective partial traces we obtain the dynamical equations
\begin{align} \label{eq:2L_EFG_dyn_eq}
\begin{split}
    \rmi\partial_t\rho_{2L} &\approx \com{H_{2L} + H_E(t)}{\rho_{2L}} + \rmi\Gamma\mathcal{L}[\sigma]\rho_{2L} + \mathcal{O}\left(\norm{Q}\right) \\
    \L Q_{\mu\nu} \R (t) &= \tr_{2L}{\left\{\rho_{2L}(t) Q_{\mu\nu}\right\}} + \mathcal{O}\left(\norm{Q}^2\right) \\
    \rmi\partial_t\rho_{S} &= \com{H_B + \L Q_{\mu\nu} \R (t) I_\mu I_\nu}{\rho_S},
\end{split}
\end{align}
where we are denoting the density operator of the electronic two-level system and of the spin system as  
$\rho_{2L}$ and $\rho_{S}$, respectively. The interaction of the open two-level system with an external electric field will lead to damped oscillations (see Section~\ref{sec:2L_sys} and panels (a) and (b) of Figure~\ref{fig:2level_EFG_pulse}), asymptotically approaching the steady-state solution~\eqref{eq:2level_oo} with a decay rate of $\Gamma$. If the external electric field is turned off, the two-level system will decay to its ground state. The key feature of our proposed protocol is now the following: Enforcing a periodic repetition of this process via a pulsed excitation, the nuclear spin system can be controlled by its quadrupole-mediated interaction with the electronic two-level system, which features different EFG tensor values in different electronic states. A graphical illustration of the proposed scheme is given in Figure~\ref{fig:2level_EFG_scheme}, where the pulse duration $\tau$, the frequency of the external undetuned field $\omega\approx\omega_0$, the decay rate $\Gamma$, as well as the EFG tensors of ground and excited state are illustrated. This is the basic principle of ONER.

Formally, we introduce a pulsed external electric field, corresponding to a cosine modulated with a square pulse, i.e. $E(t) = E_0\,\Theta(t\,\mathrm{mod}\,\tau \in (0,\tau/2)) \cos(\omega t)$, with $t\,\mathrm{mod}\,\tau$ denoting the fraction of $t$ that is in an interval $\left(n\tau,(n+1)\tau\right)$ for some $n\in\mathbb{N}$. This way, a pulsed modulation of the population of the excited state can be achieved, as illustrated in the panels (a) and (b) of Figure~\ref{fig:2level_EFG_pulse}. It shows one period of the square pulse enveloping function of $E(t)$ (red dashed line) and the resulting population of the excited state of the two-level system (blue solid line). The repetition rate $\tau^{-1}$ of the square pulse is chosen such that it matches the transition energy of the spin system for a specific transition. Note that $\frac{1}{\tau} \ll \Omega = \mathcal{O}(\mathrm{GHz})$, since the Zeeman energy splitting of the spin system is of the order of $\mathrm{MHz}$. As long as~\eqref{eq:2L_condition} is satisfied, the general analysis does not change since $\exp\left(-\Gamma\tau / 2\right)\ll 1$. This means that the excited state fully decays into the ground state after half a period of the pulse sequence. This way, a periodically pulsed modulation of the excited state population is achieved, which translates into a corresponding modulation of the EFG tensor and therefore also the NQI tensor quantities. For a sufficiently large decay rate, the steady-state solution is reached quickly, and the population of the excited state resembles a pulsed square function in good approximation. For lower decay rates, more Rabi oscillations appear before a steady-state is reached and the edge decays more slowly after the signal has been turned off, as it is illustrated in Figure~\ref{fig:2level_EFG_pulse}.

In any case, the density matrix of the two-level system is periodically modulated with period $\tau$. For convenience, the Fourier coefficients of the excited state population are compared in panels (c) and (d) of Figure~\ref{fig:2level_EFG_pulse}. The zeroth and first Fourier component are well approximated by the Fourier components of a square pulse, i.e. the steady-state solution.

From equation~\eqref{eq:2L_EFG_dyn_eq} it follows that the time dependence of the quadrupole interaction is directly inherited from the time dependence of the two-level density matrix. This can be used to determine the effective energy splitting of the spin system and the corresponding Rabi frequencies for specific transitions. In order to obtain analytical results of these quantities we investigate the effective quadrupole coupling tensor $\L Q \R$. Since the quadrupole interaction tensor is real and symmetric, and the two-level density operator is hermitian with trace one, we get
\begin{align} \label{eq:2L_EFG_ee_gg_eg}
\begin{split}
    \L Q \R (t) 
    &= \tr_{2L}{\left\{\rho_{2L}(t) Q\right\}} \\
    &= \rho_{2L,ee}(t)\b{e}Q\k{e} + \left(1-\rho_{2L,ee}(t)\right)\b{g}Q\k{g} \\
    &\qquad+ 2\mathrm{Re}\left\{ \rho_{2L,eg}(t) \b{e}Q\k{g} \right\}.
\end{split}
\end{align}
Note that the off-diagonal elements of the quadrupole coupling tensor, i.e. $\b{e}Q\k{g}$ and $\b{g}Q\k{e}$, can be chosen to be real and can therefore be pulled out of the real part in the last term. Furthermore, as can be seen from~\eqref{eq:2level_oo}, the off-diagonal steady-state components of the two-level density operator in the rotating frame, $\rho_{2L,eg}'$ and $\rho_{2L,ge}'$, are (mainly) imaginary and thus cancel effectively when taking the real part and time average over a relevant timescale of the spin system. The time average vanishes since the off-diagonal elements of the density matrix in the non-rotating frame can be written as $\rho_{2L,eg}(t) = \rho_{2L,eg}' e^{-\rmi\omega t}$, with $\omega \approx \omega_0$ the frequency of the driving field. Thus, on the timescale of the spin system, we find
\begin{equation}
\mathrm{Re}{\left\{\overline{\rho_{2L,eg}}(t)\right\}} = \mathrm{Re}{\left\{\frac{1}{T} \int_t^{t+T} \rho_{2L,eg}(t')\;\d t'\right\}} \approx 0,
\end{equation}
where $\frac{2\pi}{\omega} \ll T \ll \tau$ is some averaging time. This is even more the case if dephasing terms are added for the two-level system, since the off-diagonal density matrix elements will decay even faster. This implies that only the diagonal NQI tensor components, i.e. the quadrupole coupling tensor of the ground and excited state, are relevant for the time-dependence of the effective quadrupole coupling in the spin system. Both can be expressed via the excited state population of the two-level system, as is immediately clear from~\eqref{eq:2L_EFG_ee_gg_eg} by neglecting the last term.

In order to extract the relevant time dependency of the excited state population we investigate its Fourier expansion, which is also illustrated in panels (c) and (d) of Figure~\ref{fig:2level_EFG_pulse}. Since the excited state population is real, we may also use an expansion in a real Fourier series, which yields 
\begin{align}
\begin{split}
    \rho_{2L,ee}(t) = \frac{a^{(0)}}{2} + \sum_n \left( b^{(n)} \sin(\omega_n t) + a^{(n)} \cos(\omega_n t) \right)
\end{split}
\end{align}
with $\omega_n = 2\pi n / \tau$ and $a^{(n)} = \frac{2}{\tau}\int_0^\tau\rho_{2L,ee} \cos(\omega_n t)\;\d t$ and $b^{(n)} = \frac{2}{\tau}\int_0^\tau\rho_{2L,ee} \sin(\omega_n t)\;\d t$. Due to the time symmetry of the square pulse the sine coefficients are clearly dominating, and the cosine part can be neglected except for a constant contribution. If the conditions \eqref{eq:2L_condition} hold, the steady-state approximation can be applied (see also panels (c) and (d) of Figure~\ref{fig:2level_EFG_pulse}) and the coefficients are well approximated by the coefficients of the step function with height $\rho_{2L,ee}^\infty$, which are given by
\begin{align}
\begin{split}
    a^{(0)} &= \rho_{2L,ee}^\infty \\
    a^{(n)} &= 0,\; \text{for } n > 0 \\
    b^{(n)} &= 
    \begin{cases}
        2\rho_{2L,ee}^\infty \,/\, \pi n, & n \text{ odd,} \\
        0, & n \text{ even.}
    \end{cases}.
\end{split}
\end{align}

Since the repetition frequency $\tau^{-1}$ of the pulse is adjusted to the transition energy of the spin system, the zeroth and first order contribution are most relevant. Higher frequencies of the Fourier decomposition of the excited state population have negligible influence on spin level transitions due to large detuning. Therefore, we may write
\begin{align}
\begin{split}
    \L Q \R(t) &\approx \rho_{2L,ee}(t) \b{e}Q\k{e} + (1-\rho_{2L,ee}(t)) \b{g}Q\k{g} \\
    &\approx \b{g}Q\k{g} + \left(\b{e}Q\k{e} - \b{g}Q\k{g}\right) \rho_{2L,ee}(t) \\
    &\approx \b{g}Q\k{g} + \Delta Q(e,g) \left( \frac{a^{(0)}}{2} + b^{(1)} \sin\left(\frac{2\pi t}{\tau}\right) \right),
\end{split}
\end{align}
with $\Delta Q(e,g) := \b{e}Q\k{e} - \b{g}Q\k{g}$, by applying the steady state approximation and neglecting high frequency contributions. For given NQI tensors in the ground state and the excited state of the two-level system, respectively, it is then easy to obtain the actual dynamics of the spin system by solving the respective evolution equation~\eqref{eq:2L_EFG_dyn_eq}. Within the steady-state approximation, the spin Hamiltonian in an external magnetic field $B = B_0 e_z$ can then be written as
\begin{align} \label{eq:ham_spin_effective}
\begin{split}
H &\approx -\gamma_n B_0 I_z + I_\mu I_\nu \, Q^{(0)}_{\mu\nu}(e,g) \\
&\qquad+ I_\mu I_\nu \, Q^{(1)}_{\mu\nu}(e,g)\,\sin\left(\frac{2\pi t}{\tau}\right),
\end{split}
\end{align}
where the constant quadrupole interaction term
\begin{equation}\label{eq:Q0}
    Q^{(0)}(e,g) = \b{g}Q\k{g} + \frac{\rho^\infty_{2L,ee}}{2}\Delta Q(e,g), 
\end{equation}
determines the quadrupole energy splitting (see Section~\ref{sec:quad_ham_perturb} \eqref{eq:trans_energy_corr}) and the harmonically modulated quadrupole tensor
\begin{equation}\label{eq:Q1}
    Q^{(1)}(e,g) = \frac{2\rho^\infty_{2L,ee}}{\pi}\Delta Q(e,g),
\end{equation}
determines the magnitude of the Rabi frequency of the spin system (see Section~\ref{sec:trans_el} \eqref{eq:trans1} and \eqref{eq:trans2}). The EFG tensor of the ground state and the excited state of the two-level system can be obtained via ab initio methods. The repetition rate (i.e. the energy splitting between spin states of interest) and the resulting Rabi frequency of the spin level transitions can then be calculated analogously to Section~\ref{sec:quad_ham_perturb} and Section~\ref{sec:trans_el}, respectively.

Note that it is important to include the quadrupole splitting which stems from $Q^{(0)}$ into the calculation of the repetition rate $\tau^{-1}$ in order to avoid detuning. Furthermore, the quadrupole splitting and the spin-level Rabi frequency only depend on $\Omega / \Gamma$ via the steady-state population of the excited state $\rho_{2L,ee}^\infty$, if the conditions \eqref{eq:2L_condition} are satisfied. Thus, the intensity of the laser field should be chosen such that the Rabi frequency of the two-level system lies in the range of GHz. For a dipole moment of approximately 1 Debye, this corresponds to an electric field amplitude in the order of $10^5\;\mathrm{\frac{V}{m}}$.

\subsubsection{Numerical simulation for $^9$Be}
We pick a single $^9$Be atom for a first numerical simulation of ONER, using the Molpro program package~\cite{Molpro,Molpro1,Molpro2} to calculate the relevant parameters, i.e. the $^1P^o\leftarrow{}^1S$ transition energy as well as the electric field gradient $\Phi_{\mu\nu}$ of the both states as a function of an applied external, electric field. The choice of beryllium is motivated by the fact that it features optical excitations in the UV/visible regime and a singlet electronic ground state; the total electron spin is zero and hyperfine coupling effects, most noteworthy the substantial contributions stemming from the Fermi contact term for s orbitals, are not relevant. Employing the aug-cc-pVTZ basis set~\cite{Kendall1992}, we combine a multiconfigurational self-consistent field calculation (MCSCF~\cite{Knowles1985, Werner1985}) with a follow-up multireference configuration interaction approach (MRCI~\cite{WK88, KW92}) to account for dynamic correlation. For computational efficiency, the atom is treated within the C$_{2v}$ molecular symmetry group. A symmetrically balanced active space involving the orbitals 6/3/3/0 has been chosen with respect to the internal ordering A$_1$/B$_1$/B$_2$/A$_2$. With this setup, a perfect degeneracy of the three sublevels of the P state is preserved at zero electric field, and an excellent excitation energy of 5.30~eV is obtained for the $^1$P$^o\leftarrow{}^1$S transition, which deviates from the experimental value of 5.28~eV tabulated at NIST by less than 0.5 percent~\cite{NIST_ASD}. An evaluation of the electric field gradient of Be in its well-studied $^3$P$^o$ lowest triplet state, calculated with the same settings, produces a value of -0.1199~a.u., which agrees well (about 4\% deviation) with earlier calculated values from literature~\cite{Sundholm1991}. Sternheimer shielding effects lie within this uncertainty~\cite{Sinanoglu1973}. Note that a single value is sufficient to characterize the EFG tensor in this case due to spherical symmetry~\cite{Fowler1989}.

\begin{figure}[t]
    \centering
    \includegraphics[width=0.25\textwidth]{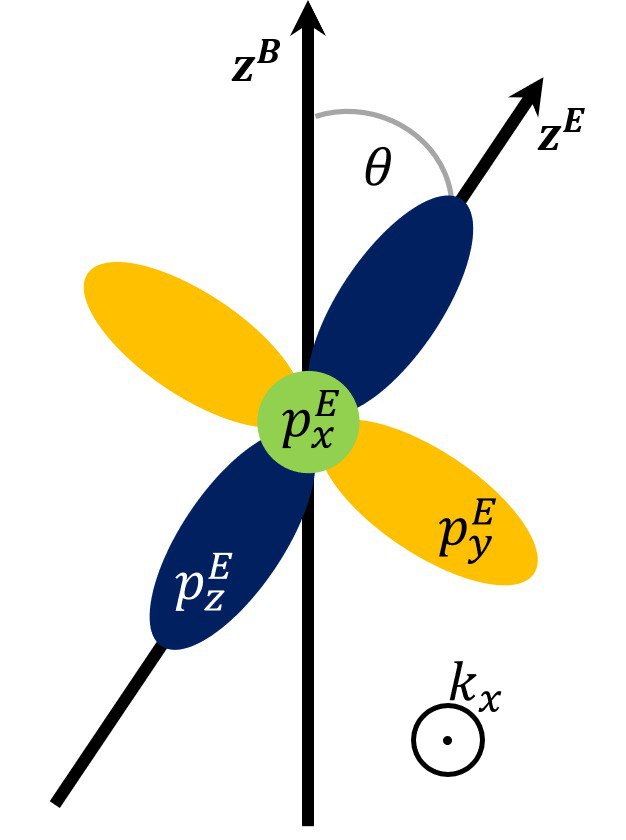}
    \caption{Orientations of the external magnetic field $B$, the electric field $E$, the propagation direction of the pulsed signal $k_x$, and the three sublevels ($p_x^E$, $p_y^E$, $p_z^E$) of the $^1$P$^o$ electronically excited state of the $^9$Be atom. The angle between the $z^B$-axis and the $z^E$-axis is denoted as $\theta$.}
    \label{fig:Li_setup}
\end{figure}

$^9$Be has a non-zero scalar quadrupole moment of $0.0529(4)$~barn \cite{tabel_of_quadrupole_moments,stone2005table}. Its gyromagnetic moment is given by $\gamma_n^{\mathrm{Be9}} = -1.17749(2)\;\mu_N \approx 8.9755\;\mathrm{\frac{MHz}{T}}$ as given in Ref.~\cite{stone2005table}, where $\mu_N = 7.622593285(47)\;\mathrm{\frac{MHz}{T}}$ is the nuclear magneton. $^9$Be has a total nuclear spin quantum number of $I=3/2$, leading to an energy scale of several $\mathrm{MHz}$ for the Zeeman Hamiltonian of the spin system. We assume that the $^9$Be atom is subject to a constant magnetic field $B$. The direction of the latter will be used to define a reference axis of the spin system, as the Zeeman splitting is dominant for the nuclear spin. Also, a constant electric field $E$ is applied to set a reference axis for the two-level system and to tune the Rabi frequency of the spin transitions. An illustration of the coordinate frames is given in Figure~\ref{fig:Li_setup}. Quantities which are frame-dependent will be marked with a superscript $E$ or $B$ for the coordinate frame being either aligned with the electric or the magnetic field, respectively. Both frames have the same $x$-axis, which we choose to be the direction of propagation of the laser field. The angle between the $z^B$-axis and the $z^E$-axis is denoted as $\theta$.

\begin{figure}[t]
    \centering
    \includegraphics[width=0.49\textwidth]{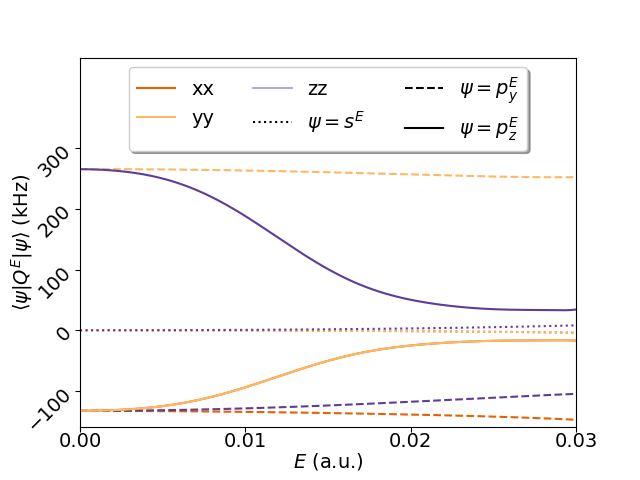}
    \caption{Non-zero components ($xx$, $yy$, $zz$) of the nuclear quadrupole interaction (NQI) tensor for $^9$Be are shown for a laboratory frame aligned with the electric field. The $s^E$-ground state NQI $\b{s^E}Q^E\k{s^E}$ is shown with dotted lines, the $p^E_z$-excited state NQI $\b{p^E_z}Q^E\k{p^E_z}$ is shown with a solid line and the $p^E_y$-excited state NQI components $\b{p^E_y}Q^E\k{p^E_y}$ are shown with dashed lines. For the $s^E$-ground NQI and the $p^E_z$-excited state NQI  the $xx$ and the $yy$ component overlap, due to symmetry.  At zero field the $s^E$-ground state is spherical symmetric and thus has vanishing NQI. Also for finite field strength the NQI of the $s^E$-ground state is negligibly small. Both $p^E_z$-excited state and $p^E_y$-excited state have similar NQI for zero field, but with swapped axis. For non-zero field the behavior differs, since the field is aligned with the $z^E$-axis.}
    \label{fig:efg_eg}
\end{figure}

In Figure~\ref{fig:efg_eg}, the non-zero components of the NQI tensor are shown in the $E$-frame. The energy splitting of the NQI lies in the range of several $\mathrm{kHz}$, which justifies the treatment of the energy correction via perturbation theory as discussed in Section~\ref{sec:quad_ham_perturb}. 

In more detail, Figure~\ref{fig:efg_eg} compares the electric field-dependent NQI tensors for $s^E$-ground state $\b{s^E}Q^E\k{s^E}$, the $p^E_y$-excited state $\b{p^E_y}Q^E\k{p^E_y}$, and the $p^E_z$-excited state $\b{p^E_z}Q^E\k{p^E_z}$. Note that the NQI for the $p^E_x$-excited state is identical with $\b{p^E_y}Q^E\k{p^E_y}$ but with $xx$ and $yy$ components swapped. The shapes of the respective NQI tensors of ground state and excited state, shown in Figure~\ref{fig:efg_eg}, are crucial for the calculation of the repetition rate and the obtained Rabi frequency of the spin level transitions. Due to spherical symmetry of the $s^E$-ground state it has a vanishing NQI at zero field. Also, for a finite field strength, the respective NQI lies in the range of a few kHz and is thus negligibly small compared to the NQI of the excited states. The $p^E$-excited states have a non-vanishing quadrupole interaction of the same magnitude at zero field but with interchanged axis. This is the expected behavior, since p-orbitals have a non-vanishing EFG tensor at the origin. The different behavior of $p^E_z$-excited state and $p^E_y$-excited state is caused by the constant electric field in $z^E$-direction.

\begin{figure}[t]
    \centering
    \includegraphics[width=0.49\textwidth]{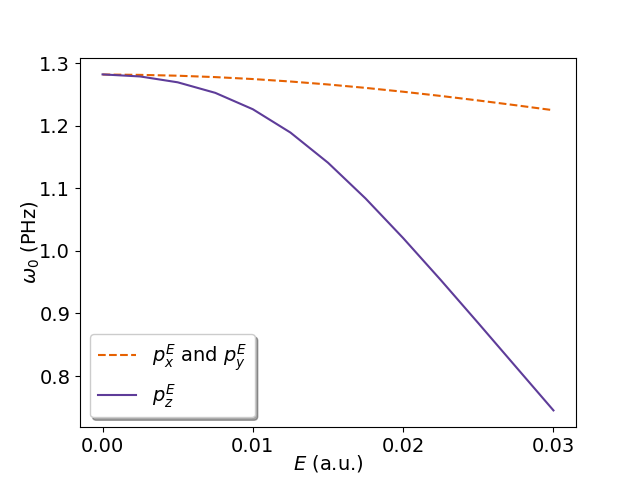}
    \caption{Stark splitting of the $^1P^o\leftarrow{}^2S$ transition of $^9$Be in an external electric field. $p^E_x$- and $p^E_y$-transitions remain degenerate, while the $p^E_z$-transition occurs at a lower energy.}
    \label{fig:Li_gs_pxyz}
\end{figure}

The corresponding electronic excitation energies of the $^1P^o\leftarrow{}^1S$ transition are illustrated in Figure~\ref{fig:Li_gs_pxyz}. As expected for a Stark splitting in the external field, the $p^E_x$- and $p^E_y$-transitions remain degenerate, while the $p^E_z$-transition occurs at a different energy.
% 1.28~PHz, 234~nm, 5.30~eV
The excitation energy for this two-level system lies around 1.28~PHz, which corresponds to 234~nm or 5.30~eV. We assume no detuning, i.e. $\omega\approx\omega_0$, and choose the amplitude of the electric field such that a Rabi frequency in the range of GHz is obtained for the two-level system. For a dipole moment of approximately 1 Debye this corresponds to an electric field amplitude of the order of $10^5\;\mathrm{\frac{V}{m}}$. The Rabi frequency of the two-level system should be chosen such that the damping $\Gamma$ is approximately of the same order; for the further discussion we choose $0.4\;\Omega$. This ensures that the steady-state approximation is valid and leads to a steady-state population of
\begin{equation}
    \rho_{2L,ee}^{\infty} = \frac{25}{54},
\end{equation}
as obtained via~\eqref{eq:2level_oo} with $\gamma_\perp = \Gamma / 2$. With the parameter values set as motivated above, the steady-state approximation holds and we can apply the analysis of the previous section.

\begin{figure}[htb!]
    \centering
    \includegraphics[width=0.49\textwidth]{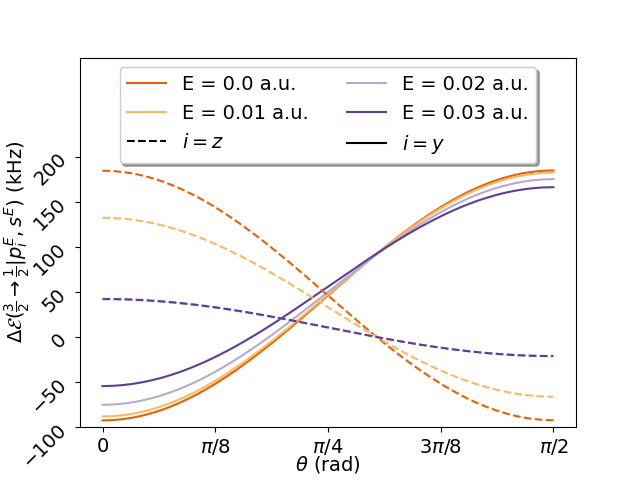}
    \caption{Quadrupole energy corrections in kHz for different electric field strengths $E$ and different angles $\theta$ between $z^E$ and $z^B$ for electronic excitations into the $p_y^E$-state (solid line) and $p_z^E$-state (dashed line), respectively. The corrections are shown for the $3/2 \leftrightarrow 1/2$ transition, but are identical to those for the $3/2 \leftrightarrow -1/2$ transition. The corrections for $-1/2 \leftrightarrow -3/2$ and the $1/2 \leftrightarrow -3/2$ only differ in sign. All quadrupole energy corrections are in the order of 100 kHz. The calculations were performed for a two-level steady-state population of $\rho_{2L,ee}^{\infty} = \frac{25}{54}$, which corresponds to $\Gamma = 0.4 \Omega$.}
    \label{fig:final_energy_splitting}
\end{figure}

From the NQI tensor components shown in Figure~\ref{fig:efg_eg} we can calculate the constant NQI tensor $Q^{(0),E}(p_i^E,s^E)$ via \eqref{eq:Q0} and the harmonically modulated NQI tensor $Q^{(1),E}(p_i^E,s^E)$ from \eqref{eq:Q1} in the $E$-frame for $i\in\{x,y,z\}$. Since the laboratory frame of the spin system is the $B$-frame we have to apply a rotation to the respective NQI tensors. The corresponding matrix for a rotation around the common $x$-axis by an angle $\theta$ is given by
\begin{align}
    R(\theta) = 
    \begin{bmatrix}
        1 & 0 & 0 \\
        0 & \cos(\theta) & -\sin(\theta)\\
        0 & \sin(\theta) & \cos(\theta)
    \end{bmatrix}.
\end{align}
The respective NQI tensors in the $B$-frame are then given by
\begin{align}
\begin{split}
    Q^{(j),B}(p^E_i,s^E) &= R(\theta) \; Q^{(j),E}(p^E_i,s^E)\; R(\theta)^\top,
\end{split}
\end{align}
for $j\in\{0,1\}$ and $i\in\{x,y,z\}$. Note that the $E$-frame still remains the reference for electronic states. The NQI tensor values in the $B$-frame can now be used to calculate the quadrupole energy correction and thus the correct repetition rate of the laser pulse, the $\Delta m_I=\pm1$ and $\Delta m_I=\pm2$ transition elements, and the corresponding Rabi frequency of the spin level transitions. In Figure~\ref{fig:final_energy_splitting}, the transition energy corrections for the different spin level transitions are given in units of kHz for different electric field strengths $E$ and angles $\theta$ between $z^E$ and $z^B$. Only the quadrupole corrections are shown, i.e. equation~\eqref{eq:trans_energy_corr} without the Zeeman splitting. The magnitude of the Zeeman splitting depends on the magnitude of the external magnetic field. However, we assume that the external field is in the order of Tesla, leading to a MHz Zeeman splitting. We denote the quadrupole energy correction for the spin level transition from state $\k{m_I}$ to $\k{m_I'}$ for given electronic excitation as $\Delta\mathcal{E}^B(m_I \rightarrow m_I' \vert p^E_i,s^E)$ with $i\in\{x,y,z\}$. Figure~\ref{fig:final_energy_splitting} shows the transition energy corrections for the $3/2 \leftrightarrow 1/2$ transition, which are identical to those for the  $3/2 \leftrightarrow -1/2$ transition.  The correction energies for the $-1/2 \leftrightarrow -3/2$ transition and the $1/2 \leftrightarrow -3/2$ only differ in sign. The $1/2 \leftrightarrow -1/2$ transition is not shown since it has a vanishing transition element, as can be seen from~\eqref{eq:trans1}. The quadrupole correction energy for the $p_z^E$-excited state NQI shows stronger changes with increasing electric field strength of the constant field $E$, while the quadrupole correction for the $p_y^E$-excited state NQI differs much less. This is reasonable, since the electric field is aligned with the $z^E$ axis.

In total, this leads to a repetition rate $\tau^{-1}$ of the pulse sequence in the range of several MHz. The quadrupole corrections are in the order of 100~kHz, and the repetition rate has to be chosen such that it matches the corrected transition energy of a spin level transition. The quadrupole correction also enables individual addressability of the spin transitions, since equidistant Zeeman transition energies are shifted individually by NQI. Given a suitable repetition rate, we observe Rabi oscillations of the spin level transitions. The Rabi frequency of the spin level transitions is determined by the modulus of the respective transition element given in \eqref{eq:trans1} and \eqref{eq:trans2} for the corresponding harmonically modulated NQI tensor $Q^{(1),B}(p^E_i,s^E)$. We denote the respective transition elements from state $\k{m_I}$ to state $\k{m_I'}$ for given electronic excitation by $g^B(m_I \rightarrow m_I' \vert p^E_i,s^E)$, with $i\in\{x,y,z\}$.

For a nuclear spin quantum number of $I = 3/2$ the prefactors $\alpha$ and $\beta$ of the $\Delta m=\pm1$ and $\Delta m = \pm2$ transitions in  \eqref{eq:trans1} and \eqref{eq:trans2}), respectively, are given by
\begin{align}
\begin{split}
    \beta_{3/2 \leftrightarrow -1/2} &= \beta_{1/2 \leftrightarrow -3/2} = \sqrt{3} / 2\\
    \alpha_{3/2 \leftrightarrow 1/2} &= \alpha_{-1/2 \leftrightarrow -3/2} = \sqrt{3} \\
    \alpha_{1/2 \leftrightarrow -1/2} &= 0. \\
\end{split}
\end{align}
The $1/2 \leftrightarrow -1/2$ transition is forbidden, whereas the other transitions might have a non-vanishing transition amplitude, depending on the respective NQI tensor.

\begin{figure*}[htb!]
    \centering
    \begin{subfigure}[b]{0.49\textwidth}
        \centering
        \includegraphics[width=0.9\textwidth]{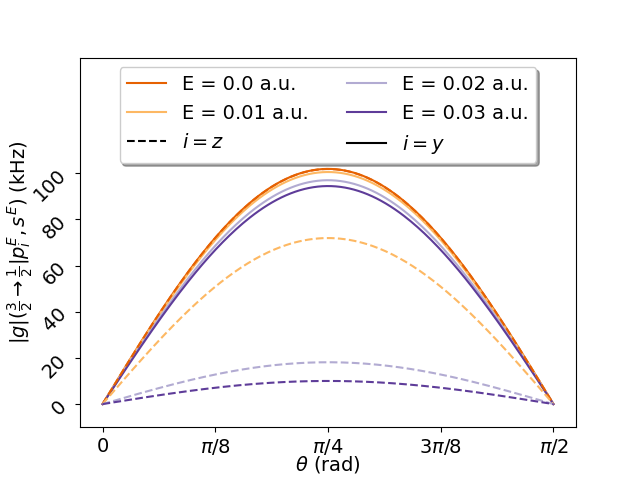}
        \caption{}
    \end{subfigure}
    \begin{subfigure}[b]{0.49\textwidth}
        \centering
        \includegraphics[width=0.9\textwidth]{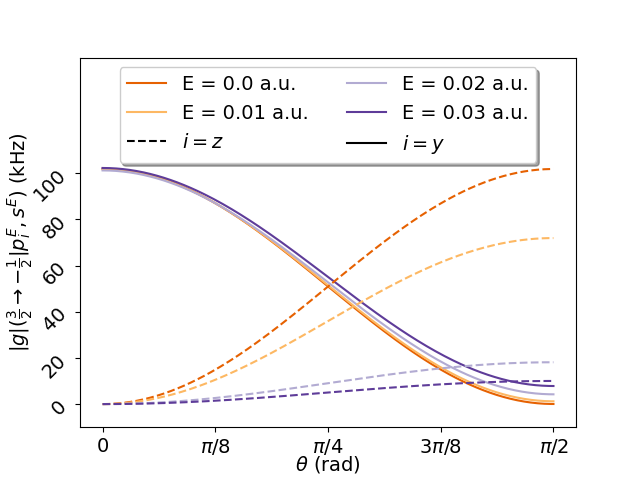}
        \caption{}
    \end{subfigure}
    \caption{Rabi frequencies in kHz for different electric field strengths $E$ and different angles $\theta$ between $z^E$ and $z^B$ for $p_y^E$-excited state (solid line) and for $p_z^E$-excited state (dashed line) NQI. Panel (a) shows the Rabi frequencies for the $\Delta m_I=\pm1$ transitions, whereas panel (b) is showing the Rabi frequencies for the $\Delta m_I=\pm2$ transitions. In both cases the Rabi frequencies are in the order of up to 100 kHz. The calculations were performed for a two-level steady-state population of $\rho_{2L,ee}^{\infty} = \frac{25}{54}$, which corresponds to $\Gamma = 0.4 \Omega$.}
    \label{fig:final_rabi}
\end{figure*}

Figure~\ref{fig:final_rabi} illustrates the dependence of the resulting Rabi frequencies on the relative angle $\theta$ between $E$-frame and $B$-frame for different values of the electric field strength $E$. Panel (a) shows the Rabi frequency for the $3/2 \leftrightarrow 1/2$ transition, which is the same for the  $-1/2 \leftrightarrow -3/2$ transition, and panel(b) shows the Rabi frequency for the $3/2 \leftrightarrow -1/2$ transition, which is also the same for the $1/2 \leftrightarrow -3/2$ transition.

In general, the Rabi frequency for the NQI tensor of the $p_z^E$-excited state can be more easily controlled with the external constant electric field. The Rabi frequency for the NQI tensor of the $p_y^E$-excited state varies much less; for the $\Delta m_I=\pm1$ transitions in particular, almost no dependence can be observed. Note that, for zero field, the transition elements of the NQI tensor for both excited states coincide.

The general shape of the $\theta$-dependence of the Rabi frequency can be explained as follows. For an angle of $\theta = 0$ the NQI of both excited states is diagonal and therefore no $\Delta m_I=\pm1$ transitions can be driven as the transition element vanishes. This is immediate from~\eqref{eq:trans1}. The same happens for $\theta = \pi/2$. For $\theta = \pi/4$, the off-diagonal elements are maximized, as can be easily derived from the shape of the rotation matrix. 

For the $p_z^E$-excited state also $\Delta m_I=\pm2$ transitions cannot be driven at $\theta = 0$, since $xx$ and $yy$ components coincide (see \eqref{eq:trans2}). For the $p_y^E$-excited state, the $xx$ and $yy$ components of the corresponding NQI tensor are maximally different in this case, leading to a maximum of the transition element at $\theta = 0$. For the NQI of the $p_z^E$-excited state, the $xx$ and $yy$ components are maximally different for an angle of $\pi / 2$, since the corresponding rotation into the $B$-frame swaps the $yy$ and $zz$ component. Depending on the transitions one is interested in, the angle has to be chosen accordingly. A reasonable choice would be $\theta = \pi / 4$, since all transitions can then be addressed. The resulting Rabi frequencies lie in the range of 100~kHz, depending on the orientation and the magnitude of the constant electric field. For comparison, the Rabi frequency in the NER studies of Ref.~\cite{original_nature} was found at 68~kHz.

\begin{figure*}[htb!]
    \centering
    \begin{subfigure}[b]{0.49\textwidth}
        \centering
        \includegraphics[width=0.9\textwidth]{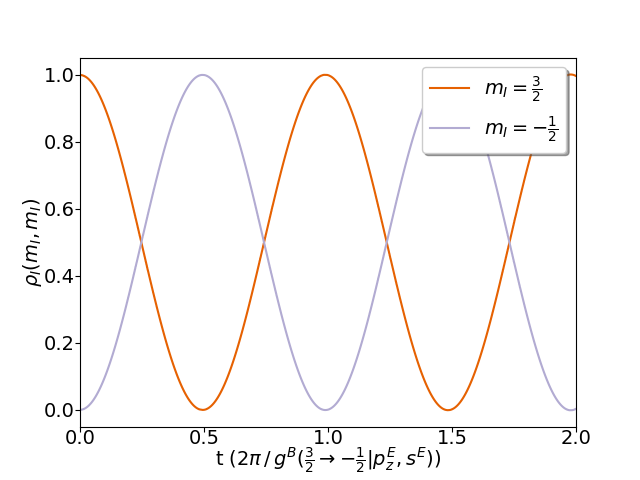}
        \caption{}
    \end{subfigure}
    \begin{subfigure}[b]{0.49\textwidth}
        \centering
        \includegraphics[width=0.9\textwidth]{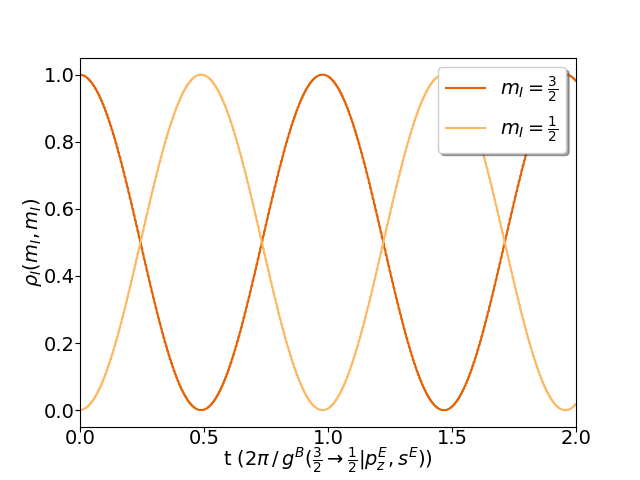}
        \caption{}
    \end{subfigure}
    \caption{Rabi oscillations of the numerical simulation of a quadrupolar nucleus with nuclear spin $I=3/2$ via pulsed excitation of a coupled two-level system subject to an external constant magnetic field of magnitude 1 Tesla. The simulation parameters are chosen for an electric field strength of $E = 0.01$ a.u. with an angle of $\theta = \pi/4$ to the magnetic field axis. The two-level system parameters were chosen as $\Gamma = 0.4 \Omega$ and $\Delta = 0$ with $\Omega = 1$ GHz. Electronic excitation was performed into the $p_z^E$-excited state. The $y$-axis shows the population of the respective spin states $m_I$. Panel (a) shows the $\frac{3}{2} \leftrightarrow -\frac{1}{2}$ transition, panel (b) the $\frac{3}{2} \leftrightarrow \frac{1}{2}$ transition. The $x$-axes are normalized to the analytically calculated Rabi frequency. It can be seen that in both cases the analytical frequency and the numerical are in good agreement. Slight deviations can be seen, since maxima and minima are not located at integer and half-integer values of the normalized time.}
    \label{fig:plt_transition_pulse}
\end{figure*}

Finally, in Figure~\ref{fig:plt_transition_pulse}, we show the resulting Rabi oscillations from a numerical simulation of the coupled system, i.e. a numerical simulation of the open two-level system coupled to a spin system via nuclear quadrupole interaction. The total Hamiltonian from \eqref{sec:nuc_quad_ham} with decay operators for the two-level system is simulated for an external pulsed excitation with repetition rate $\tau^{-1}$ with the standard Master equation solver of the Python library QuTiP~\cite{qutip1,qutip2}. The simulation parameters are chosen as $\Gamma = 0.4 \Omega$ and $\Delta = 0$, with $\Omega = 1$ GHz, similar to the above analytical investigation of the system. The angle between electric field and magnetic field is chosen as $\theta = \pi/4$, the constant electric field strength was chosen to be $E = 0.01$ a.u. and the magnetic field strength was set to 1~Tesla. An electronic excitation into the $p_z^E$-excited state is assumed. The repetition rate is chosen such that it matches the desired spin level transition energy including the quadrupole correction term. Note that the analytical treatment is necessary for this step,  as it delivers the corrected repetition rate of the laser pulses. Figure~\ref{fig:plt_transition_pulse} shows the Rabi oscillations of the numerical simulation of the spin system for different pulse frequencies matching the transition energies of the spin system for a $3/2\leftrightarrow 1/2$ and a $3/2 \leftrightarrow -1/2$ transition, as derived from the analytical considerations. The time axis is normalized to the analytically obtained Rabi frequency of the respective transition. It can be seen that there is a slight deviation of the numerically obtained Rabi frequency and the analytical one, since the maxima and minima are not perfectly at integer and half-integer values. However, they are in good agreement.

To summarize, this shows that a nuclear spin system can be controlled by pulsed electronic excitation of a coupled two-level system. Since the transitions appear at different energies, they can be addressed individually by selecting an appropriate repetition rate for the external pulses. An additional decay channel of the two-level system, stemming from the interaction with the spin system, is not problematic since the decay of the two-level system is mainly governed by external couplings. In fact, the two-level system can be considered effectively independent of the spin system, which enables a simplified analysis of the combined system and the imposed quadrupole interaction between electronic excitation and nuclear spin manipulation. The simplified analytic discussion is corroborated by the numerical results obtained for the total coupled system.

\section{Conclusion}
\label{sec:conclusion}
Fundamental principles and practical implications of nuclear quadrupole resonance as a tool for quantum computing were investigated. A detailed theoretical description of nuclear quadrupole coupling was developed from first principles. The quadrupole interaction Hamiltonian was derived from the molecular Hamiltonian in a consistent manner by employing Taylor expansions for non-pointlike nuclei. Important time-independent and time-dependent properties of the nuclear quadrupole Hamiltonian were discussed, and the coupling of electric field gradient and nuclear spin was investigated in detail. Within the adiabatic approximation, and assuming a quasi-independence of the electronic system from the nuclear spin system, a consistent, general description for future discussions of NER and NAR experiments was obtained (a summary of the applied approximations can be found in Appendix~\ref{sec:A6}). Our formalism exceeds simpler phenomenological models and lays a consistent foundation for commonly applied approximations, which can be obtained as special cases of our more general description (a typical phenomenological model is discussed in Appendix~\ref{sec:A5}).

Putting this general description to practice, we further propose a new scheme for nuclear electric resonance using pulsed electronic excitation. Polarized laser pulses in the UV/visible spectrum can be used to drive spin transitions via a coupling of the EFG tensors in different electronic states of an atomic or molecular system to the nuclear quadrupole moment. We refer to this technique as `optical nuclear electric resonance' (ONER). It exploits changes in the electric field gradient at the position of the nucleus, which are induced by electronic excitation, e.g. from the electronic ground state into a suitable excited state. In a first numerical test on atomic $^9$Be as a benchmark, Rabi oscillations for nuclear spin transitions in the order of several kHz are predicted.

Based on these first results, we believe that ONER has the potential to link coherent nuclear spin manipulation with well-established concepts of optoelectronics and nanophotonics. As a third paradigm aside NER and NAR, this new protocol offers the advantage of an increased flexibility with respect to the addressing of molecular spin systems: It combines a selectivity which is intrinsic to electronic excitations and their specific effect on the electric field gradient at the various nuclear positions of a molecular system. Furthermore, it has the ability to select specific spin transitions of these nuclei via pulsed laser light, by tuning the repetition rate to a certain transition of interest.

Future research will be devoted to the simulation of larger, more complicated but experimentally accessible systems with the possibility to address and couple specific nuclear spins, e.g. within the same molecule or a given molecular qubit register, through different electronically excited states. Potential candidate systems are metal complexes with reduced magnetic noise and minimal vibrational coupling~\cite{GaitaArino2019}. Alternatively, regarding the individual addressing of spatially separated quantum systems such as cold atoms, ions, or solid-state qubits e.g. via light-shift gradients~\cite{Gardner1993,Thomas1995} or non-linear response in two-level systems~\cite{Gorshkov2008}, new techniques of quantum optical control may emerge from a combination of methods.

As a final comment, we would like to emphasize that the current formalism does not involve any coupling to the electron spin, which is assumed zero. In singlet systems, the only spin-spin coupling takes place between nuclei, which might be a technical advantage, and should be the concern of future publications on the subject.

%% APPENDIX %%%%%%%%%%%%%%%%%%%%%%%%%%%%%%%%%%%%%%%%%%%%%%%%%%%%%%%%%%%%%%%%%%%%%%%%%%%%%%%%%
\begin{appendix}
\section{Overview}
Appendix~\ref{sec:A1} contains a detailed derivation of the nuclear quadrupole Hamiltonian. Appendix~\ref{sec:A2} provides a brief introduction to density operators and open quantum systems. Appendix~\ref{sec:A3} gives an overview of common units and the order of energy splittings due to quadrupole effects. Appendix~\ref{sec:A4} contains a description of graphical illustrations of EFG tensors for the sake of an improved, visual understanding of the tensor components occurring in the main text. In Appendix~\ref{sec:A5} we provide a phenomenological  description of the nuclear quadrupole coupling that is commonly used to describe nuclear electric resonance or nuclear acoustic resonance. Appendix~\ref{sec:A6} summarizes the approximations made in the theoretical description of NER, NAR, and ONER.

\section{Derivation of the nuclear quadrupole Hamiltonian}\label{sec:A1}
We start from the usual many-particle Hamiltonian of molecular physics in the absence of external fields. In natural units ($\hbar = \frac{1}{4\pi\epsilon_0} = m_e = a_0 = 1$) it reads
\begin{align} \label{eq:mol_ham}
\begin{split}
H^{(0)} = 
&\overbrace{\sum_i -\frac{1}{2}\nabla_i^2}^{T_E}
\; + \; 
\overbrace{\sum_A -\frac{1}{2M_A}\nabla_A^2}^{T_N^{(0)}} 
\; + \; 
\overbrace{\sum_{i<j} \frac{1}{\abs{r^{(i)} - r^{(j)}}}}^{V_{EE}}
\;  \\ &+ \; 
\underbrace{\sum_{A<B} \frac{Z_A Z_B}{\abs{R^{(A)} - R^{(B)}}}}_{V_{NN}^{(0)}} 
\; + \; 
\underbrace{\sum_{i,A} -\frac{Z_A}{\abs{r^{(i)} - R^{(A)}}}}_{V_{EN}^{(0)}},
\end{split}
\end{align}
with lower- and upper-case indices for electron and nuclear coordinates, respectively. The position of the $i$-th electron is denoted as $r^{(i)}$, and $\nabla_i$ is the derivative with respect to this position. The position of the $A$-th nucleus is denoted as $R^{(A)}$ and $\nabla_A$ is the derivative with respect to this position. Charge and mass of the $A$-th nucleus are denoted as $Z_A$ and $M_A$, respectively. Thus, $T_E$ represents the electron kinetic energy, $T_N^{(0)}$ the nuclei kinetic energy, $V_{EE}$ the electron-electron interaction, $V_{NN}^{(0)}$ the nucleus-nucleus interaction and $V_{EN}^{(0)}$ the electron-nuclei interaction. The superscript $(0)$ will turn out convenient in the further discussion.

In this simplified Hamiltonian, the nuclei are thought of as point-like particles of mass $M_A$ and charge $Z_A$; their actual, non-trivial charge distribution is neglected. However, since the deviation of proton positions from the center-of-charge of the respective nucleus can be considered small, we can apply a Taylor series expansion up to second order to obtain correction terms for a non-point-like charge distribution of the respective nuclei. Introducing center-of-charge coordinates $R^{(A)} = \frac{1}{Z_A}\sum_{p_A} R^{(A,p_A)}$ and denoting the deviation of each proton from this center as $\delta R^{(A,p_A)} = R^{(A,p_A)} - R^{(A)}$, we obtain the corrected total Hamiltonian
\begin{align} \label{eq:full_hamiltonian_quad}
\begin{split}
H = H^{(0)} + \frac{1}{6}\sum_A \Phi_{\mu\nu}^{(A)} \mathcal{Q}_{\mu\nu}^{(A)},
\end{split}
\end{align}
with $\Phi_{\mu\nu}^{(A)}$ as the total electric field gradient tensor at the position of the $A$-th nucleus,
\begin{align} \label{eq:EFG_total}
\begin{split}
    \Phi_{\mu\nu}^{(A)}
    =\!\sum_{B \neq A}& \frac{Z_B}{\abs{\mathfrak{R}^{(A,B)}}^5}\left( 3\mathfrak{R}^{(A,B)}_\mu \mathfrak{R}^{(A,B)}_\nu - \delta_{\mu\nu} \abs{\mathfrak{R}^{(A,B)}}^2 \right) \\
    -&\sum_i \frac{1}{\abs{\mathfrak{r}^{(i,A)}}^5} \left( 3\mathfrak{r}^{(i,A)}_\mu \mathfrak{r}^{(i,A)}_\nu - \delta_{\mu\nu} \abs{\mathfrak{r}^{(i,A)}}^2 \right),
\end{split}
\end{align}
with definitions $\mathfrak{r}^{(i,A)} := r^{(i)} - R^{(A)}$ and $\mathfrak{R}^{(A,B)} := R^{(A)} - R^{(B)}$ for electron-nucleus and nucleus-nucleus difference vectors, respectively. $\mathcal{Q}_{nm}^{(A)}$ denotes the quadrupole moment of the respective nucleus, which can be expressed as
\begin{align}\label{eq:quadrupol_position_vector}
\mathcal{Q}_{\mu\nu}^{(A)} = \sum_{p_A} \left(3\delta R^{(A,p_A)}_\mu \delta R^{(A,p_A)}_\nu - \delta_{\mu\nu} \abs{\delta R^{(A,p_A)}}^2\right).
\end{align}
The nuclear quadrupole moment of each nucleus can be related to the total nuclear spin by employing the Wigner-Eckart theorem. The nuclear quadrupole tensor is a traceless symmetric second rank tensor. It is proportional to the symmetric traceless second rank total angular momentum tensor $\frac{3}{2} \left( I_\mu I_\nu + I_\nu I_\mu \right) - \delta_{\mu\nu} I^2$ in a subspace with constant nuclear spin quantum number $I$. This is satisfied, in good approximation, for the case of nuclear electric resonance, since the orbital angular momentum of the nucleus can be considered constant. Thus, it is sufficient to calculate the matrix elements in the basis $\k{I,m_I}$ of the magnetic quantum numbers of the nuclear spin, with spin quantum number $I$ and magnetic quantum number $m_I$. Calculating the proportionality constant in the $\k{I,I}$ state yields
\begin{align}
    \mathcal{Q}_{\mu\nu} = \frac{q}{I(2I-1)} \left( \frac{3}{2} \left( I_\mu I_\nu + I_\nu I_\mu \right) - \delta_{\mu\nu} I^2 \right),
\end{align}
with $q := \b{I I} \mathcal{Q}_{33} \k{I I}$ as the scalar quadrupole moment of the nucleus. The numerical values of the scalar quadrupole moments of different nuclei are tabulated in Refs.~\cite{stone2005table,tabel_of_quadrupole_moments}. Note that only nuclei with nuclear spin $I > \frac{1}{2}$ can have a non-vanishing quadrupole tensor, as indicated by the proportionality constant.

Exploiting that the electric field gradient tensor $\Phi$ is symmetric and traceless, the nuclear quadrupole Hamiltonian for a single nucleus can be rewritten as
\begin{align}
\begin{split}
H_Q = \frac{1}{6} \Phi_{\mu\nu} \mathcal{Q}_{\mu\nu} = I_\mu Q_{\mu\nu} I_\nu,
\end{split}
\end{align}
with $Q_{\mu\nu} = \frac{q}{2I(2I-1)} \Phi_{\mu\nu}$.

\section{Open quantum systems}\label{sec:A2}
There are several different timescales involved and interactions with the environment of the molecular system need to be taken into account. This is achieved by introducing the density operator $\rho$, which has the properties
\begin{equation} \label{eq:densop_conditions}
    \rho = \rho^\dagger \geq 0, \qquad \tr{\{\rho\}} = 1,
\end{equation}
and, in the case of an isolated system, obeys the von Neumann equation
\begin{align} \label{eq:vonNeumann}
\begin{split}
    \rmi\hbar\partial_t \rho 
    &= \com{H}{\rho}.
\end{split}
\end{align}
As a hermitian operator the density operator can be expressed in its eigenbasis via $\rho = \sum_\alpha p_\alpha \k{\psi_\alpha}\b{\psi_\alpha}$. The expectation value of an observable $O$ is calculated by the trace $\tr\{\rho O\} = \sum_\alpha p_\alpha \b{\psi_\alpha}O\k{\psi_\alpha}$, highlighting the statistical nature of the density operator.

In open systems, the system of interest interacts with an environment. The total system, $SE$, consisting of the system of interest $S$ and the environment $E$, is treated via a joint density operator $\rho_{SE}$ whose dynamics is given by the von Neumann equation. Since we are only interested in the dynamics of the system $S$, we want to reduce this density matrix to a density matrix of that particular system only, such that expectation values of observables and matrix elements of the system remain the same as for the total system $SE$. This is reasonable because the dynamics of the environment is unknown in most cases and observables are only accessible for the system $S$. For that purpose, the reduced density operator $\rho_S$ is introduced as the partial trace of the total density operator,
\begin{equation}
    \rho_S = \tr_E\left\{\rho_{SE}\right\} = \sum_{m_E} \b{m_E} \rho_{SE} \k{m_E},
\end{equation}
for some basis $\k{m_E}$ of the environment Hilbert space. Due to the linearity of the trace operator, it is immediately clear that expectation values of observables of the system $O_S$ can be calculated via
\begin{align}
\begin{split}
    \L O_S \R 
    &= \tr\left\{ \rho_{SE} O_S\right\} = \tr_S\left\{O_S\,\tr_E \rho_{SE}\right\} \\ 
    &= \tr_S\left\{O_S\rho_S\right\}.
\end{split}
\end{align}
Additionally, it is easy to check that the reduced density operator fulfills the conditions for a density operator stated in \eqref{eq:densop_conditions} in the Hilbert space of the system $S$. In general, the Hamiltonian of the total system may be written as
\begin{equation}
    H = H_S \otimes I_E + I_S \otimes H_E + H_{SE},
\end{equation}
with $H_S$ and $H_E$ as operators acting exclusively on the system and the environment, respectively, and $H_{SE}$ as a coupling term. $I_E$ and $I_S$ are denoting unit operators in their corresponding Hilbert space. If the interaction between system and environment is negligible, i.e. $H_{SE} \approx 0$, then the dynamical equation for the reduced density operator is given by
\begin{equation}
    \rmi\hbar\partial_t \rho_S = \com{H_S}{\rho_S}
\end{equation}
as can be checked easily. In this case, the system can be treated as an isolated system. However, if the interaction between system and environment is not negligible, the treatment of the dynamics of the system becomes much more involved. A common ansatz to handle such cases is the Born-Markov approximation, which leads to a non-unitary evolution equation for the reduced density operator, often referred to as Linblad Master equation. A full derivation can be found in Ref.~\cite{quantum_optics_long_book}. One starts from the evolution equation for the total system
\begin{equation}
    \rmi\hbar\partial_t \rho_{SE} = \com{H}{\rho_{SE}},
\end{equation}
and the assumptions of initial separability, i.e. $\rho_{SE}(0) = \rho_S(0) \otimes \rho_E(0)$, separability during time evolution and constant environment (Born approximation), i.e. $\rho_{SE}(t) = \rho_S(t)\otimes\rho_E$, a short memory environment (Markov approximation) and a coarse grained system dynamics (secular approximation).

This leads to the Born-Markov Master equation of the reduced system
\begin{equation}
    \rmi\hbar\partial_t \rho_S = \com{H_S}{\rho_S} + \rmi\hbar\sum_\alpha k_\alpha \mathcal{L}[c_\alpha]\rho_S,
\end{equation}
with decay strengths $k_\alpha$ and the Linblad superoperator $\mathcal{L}$ defined by
\begin{equation}
    \mathcal{L}[c]\rho_S = c\rho_S c^\dagger - \frac{1}{2}\left( c^\dagger c \rho_S + \rho_S c^\dagger c \right)
\end{equation}
for a so-called collapse operator $c$. The prefactors $k_\alpha$ of the non-unitary evolution terms can be interpreted as the rate of the process described by the coupling operators, for example the strength of dissipation due to coupling to the environment. Note that a Master equation of this type is trace-preserving and ensures that the density operator is hermitian, which is important to ensure that the solution to the Lindblad Master equation is indeed a density operator.

\section{Units and order estimates} \label{sec:A3}
In this section we provide numerical estimates of typical energy scales within the context of quadrupole interaction. So far, atomic units have been used, i.e. $\hbar = \frac{1}{4\pi\epsilon_0} = m_e = a_0 = 1$. Within the SI, the EFG tensor is given in units of $\mathrm{\frac{V}{m^2}}$. The corresponding atomic units are $1\;\mathrm{au} = \frac{E_H}{e a_0^2}$, with $a_0$ denoting the Bohr length, $e$ the elementary charge and $E_H$ the energy in Hartree. The conversion factor to SI units is $1\;\mathrm{au} = 9.717 \times 10^{21}\;\mathrm{\frac{V}{m^2}}$.

Typical values for the scalar quadrupole moment of a nucleus are of the order of $10^{-3} - 1$ barn, where $1\;\mathrm{barn} = 10^{-28}\;\mathrm{m}^2$, as can be seen from the table of scalar quadrupole moments given in Refs.~\cite{tabel_of_quadrupole_moments,stone2005table}. Since $q$ is very small, the EFG necessary to generate a significant quadrupole interaction needs to be very large. Typically, only a microscopic mechanism in a crystal lattice or molecule, such as the distortion of covalent bonds in the vicinity of the nucleus, creates a significant EFG; values of the latter range from $10^{16}\;\mathrm{\frac{V}{m^2}}$ to $10^{21}\;\mathrm{\frac{V}{m^2}}$. For scalar quadrupole moments in the range of $10^{-3} - 1$ barn this yields an interaction strength of the order of $\mathrm{kHz}$ to $\mathrm{MHz}$.

Often, a magnetic field is applied as well to generate a Zeeman-splitting of the nuclear energy levels.
The magnitude of the gyromagnetic moment is of the order of $1 - 50\;\mathrm{\frac{MHz}{T}}$ as shown in Refs.~\cite{tabel_of_quadrupole_moments,stone2005table}. Assuming a magnetic field of approximately 1~T, this results in a level splitting in the $\mathrm{MHz}$ regime. In comparison to that, the gyromagnetic moment of the electron is given by approximately $28\;\mathrm{\frac{GHz}{T}}$. Hence, besides direct stimulation via radio frequency adsorption, only an acoustic phonon excitation in solids offers a possible coupling within this energy regime. These options give rise to the possibility of nuclear electric resonance (NER) or nuclear acoustic resonance (NAR) as experimental techniques that have been studied so far in recent years~\cite{original_nature, NAR1, NAR2}.

\section{Graphical illustration of EFG tensors}\label{sec:A4}
\begin{figure*}[htbp!]
    \centering
    \begin{subfigure}[b]{0.4\textwidth}
        \centering
        \includegraphics[width=\textwidth]{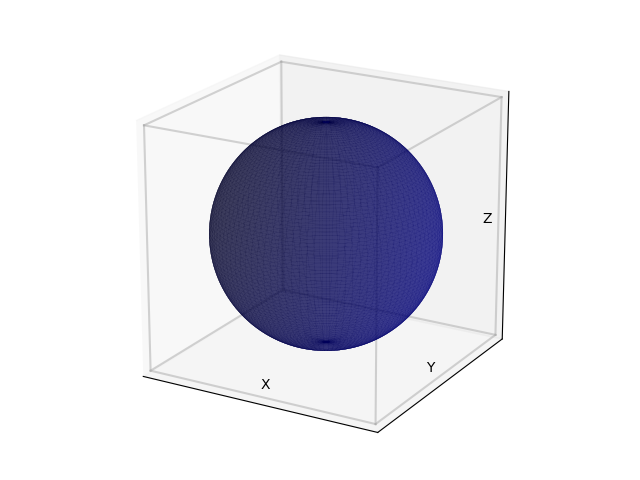}
        \caption{isotropic, $\mathrm{diag}(1,1,1)$}
    \end{subfigure}
    \begin{subfigure}[b]{0.4\textwidth}
        \centering
        \includegraphics[width=\textwidth]{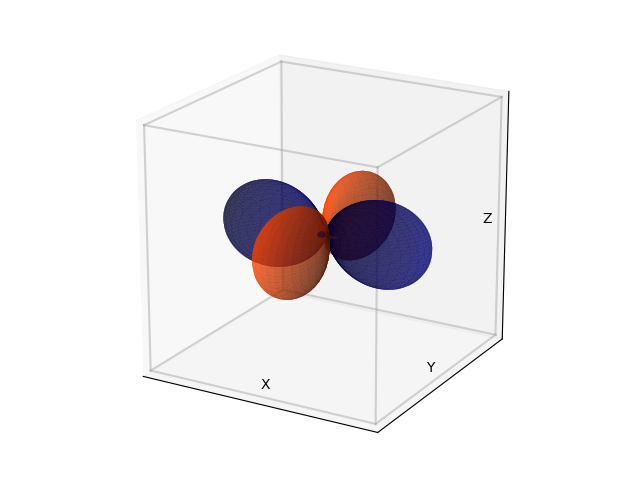}
        \caption{axial, $\eta = 1$, $\mathrm{diag}(1,-1,0)$}
    \end{subfigure}
    \begin{subfigure}[b]{0.4\textwidth}
        \centering
        \includegraphics[width=\textwidth]{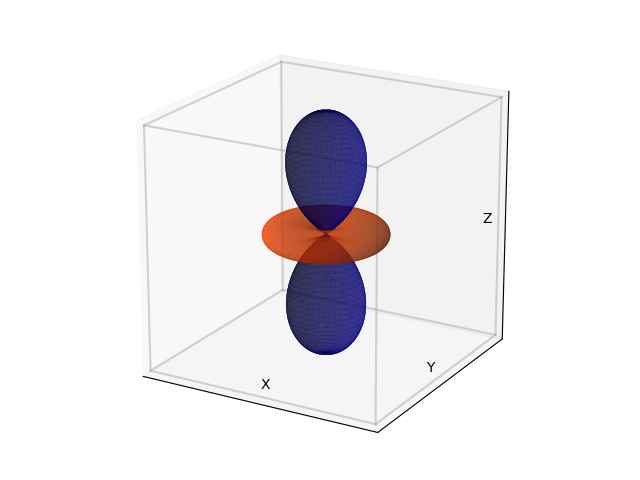}
        \caption{axial, $\eta = 0$, $\mathrm{diag}(1,1,-2)$}
    \end{subfigure}
    \begin{subfigure}[b]{0.4\textwidth}
        \centering
        \includegraphics[width=\textwidth]{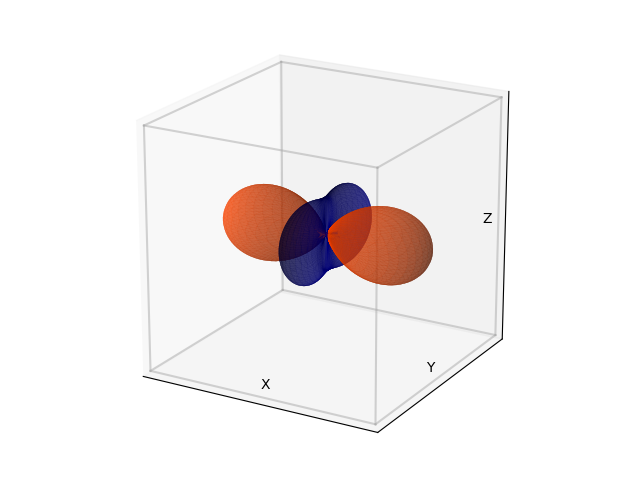}
        \caption{intermediate, $\eta = 1/3$, $\mathrm{diag}(-3,2,1)$}
    \end{subfigure}
    \caption{Graphical representation of a second rank tensor with varying asymmetry parameters $\eta$ in the principle axis system. (a) shows an isotropic second rank tensor, (b), (c) and (d) traceless second rank tensors with asymmetry parameters $\eta = 1$, $\eta = 0$ and $\eta = 1/3$, respectively.}
    \label{fig:efg_plot_simple}
\end{figure*}

In order to improve our understanding of the structure and the origin of EFG tensor components we want to display them graphically and investigate different examples of combinations of atomic orbitals, since, in the general case, linear combinations of atomic orbitals have to be considered. As defined in Ref.~\cite{efg_origins}, we will consider the function 
\begin{align}
f(r) = s\sum_{ij} r_i \L \Phi_{ij} \R r_j = s\norm{r}^2 g(\phi,\theta)
\end{align}
with $\L \Phi_{ij} \R$ being the expectation value of the EFG tensor in the state to investigate and $s$ a scaling parameter, to illustrate the EFG tensor as a three-dimensional surface plot. We set $\norm{r} = \abs{g(\phi,\theta)}$ and plot the resulting surface in blue if $g(\phi,\theta) > 0$ and orange if $g(\phi,\theta) < 0$. This is a general scheme that can be used to illustrate symmetric second rank tensors graphically. An example is given in Figure~\ref{fig:efg_plot_simple}, where the isotropic case and three feasible EFG tensors for different asymmetry parameters $\eta$ are presented. The asymmetry parameter is defined as
\begin{align}
\eta &= \frac{\abs{\Phi_{y'y'} - \Phi_{x'x'}}}{\abs{\Phi_{z'z'}}}, 
\end{align}
where the $x',y',z'$ are the eigenvectors of the EFG tensor, with $z'$ corresponding to the largest eigenvalue.
If the EFG tensor is not diagonal in the chosen laboratory frame, it will appear as a rotation of a diagonal tensor in the principle axis system to the laboratory frame. In molecular systems, the origin of the tensor is shifted to the origin of the nucleus of interest.

\section{Phenomenological description of nuclear quadrupole coupling}\label{sec:A5}
The theoretical treatment in Section~\ref{sec:NER_general} entails and justifies a commonly applied model for nuclear quadrupole coupling, which is briefly discussed here. We assume an EFG tensor created by a specific charge distribution $n$. The EFG tensor at the position of a nucleus $A$ of a molecular system (with fixed nuclei), denoted as $\Phi_{\mu\nu}^{(A)}$, is given in \eqref{eq:EFG_total}.

Since only one-electron operators appear in this expression, we can relate the expectation value $\b{\psi} \Phi_{\mu\nu}^{(A)} \k{\psi}$, for a given state $\k{\psi}$ of the system, to an integral over the total charge density
\begin{align}
\begin{split}
    n(r^{(1)}) = -N\int &\abs{\psi(r^{(1)},\dots,r^{(N)})}^2\;\mathrm{d}^3r^{(2)}\dots\mathrm{d}^3r^{(N)} \\
    &+ \sum_{B} Z_B \delta(r^{(1)} - R^{(B)}),
\end{split}
\end{align}
with $N$ as the total number of electrons in the system, and obtain
\begin{align} \label{eq:EFG_density}
    \Phi_{\mu\nu}^{(A)} = \int \frac{1}{\abs{\mathfrak{r}^{(A)}}^5} \left( 3\mathfrak{r}^{(A)}_\mu \mathfrak{r}^{(A)}_\nu - \delta_{\mu\nu} \abs{\mathfrak{r}^{(A)}}^2 \right) n(r)\;\mathrm{d}^3r,
\end{align}
again with $\mathfrak{r}^{(A)} = r - R^{(A)}$, as defined in Appendix~\ref{sec:A1}. Note that this description incorporates also movements of the nuclei, yet in a classical manner, within the Born-Oppenheimer approximation. Assuming that the position of the nucleus $A$ is fixed, which can always be done by coordinate transformation, the first term in equation~\eqref{eq:EFG_density} does not change under application of an electric field $E$ or mechanical strain $\varepsilon$ to the system. Thus, a variation of the EFG tensor is mediated through the change of the total charge density. 

In a linear expansion around the zero-point position, i.e. zero electric field and strain, we can write
\begin{align} \label{eq:EFG_dens_expand}
\begin{split}
    \Phi_{\mu\nu}^{(A)} &\approx \int\mathrm{d}^3r\; \frac{\left( 3\mathfrak{r}^{(A)}_\mu \mathfrak{r}^{(A)}_\nu - \delta_{\mu\nu} \abs{\mathfrak{r}^{(A)}}^2 \right)}{\abs{\mathfrak{r}^{(A)}}^5} \\
    &\times 
    \left( n(r)\bigg\vert_{\substack{E = 0 \\ \varepsilon = 0}} + \frac{\delta n(r)}{\delta \varepsilon_{\alpha \beta}}\bigg\vert_{\substack{E = 0 \\ \varepsilon = 0}} \varepsilon_{\alpha\beta} + \frac{\delta n(r)}{\delta E_\gamma}\bigg\vert_{\substack{E = 0 \\ \varepsilon = 0}} E_\gamma \right).
\end{split}
\end{align}
This expression suggests the functional form
\begin{align} \label{eq:EFG_pheno}
    \Phi_{\mu\nu}^{(A)} \approx \Phi_{\mu\nu}^{(A),(0)} + S_{\mu\nu\alpha\beta} \epsilon_{\alpha\beta} + R_{\mu\nu\gamma} E_\gamma
\end{align}
for the electric field gradient tensor at the position of nucleus $A$, with $\Phi_{\mu\nu}^{(A),(0)}$ denoting the EFG tensor of the unperturbed system, $S_{\mu\nu\alpha\beta}$ as a fourth rank coupling tensor of strain and the EFG tensor, and $R_{\mu\nu\gamma}$ as a third rank coupling tensor of electric field and the EFG tensor. Note that we sum over Greek indices appearing twice in one term. The tensors $S_{\mu\nu\alpha\beta}$ and $R_{\mu\nu\gamma}$ are determined by the change of the total charge density due to the interaction of the system with strain and electric field. In a simplified picture, we can deduce that the distortion of the electron hull or the atomistic environment leads to an electric field gradient at the position of the nucleus. Due to the $\sim \frac{1}{r^3}$ characteristics of the EFG tensor, we would expect that only effects in the close neighborhood of the nucleus of interest are relevant.
The coupling of an external static electric field to the EFG tensor is a well-known phenomenon, which is also known as linear quadrupole Stark effect~\cite{LQSE_R14,hyperfine_stark_effect}.

The derived relation between EFG tensor components and external quantities can be generalized to the time-dependent case, if we assume that the density relaxes instantaneously to the ground state, i.e. that the timescale of the external field variations is much larger than the timescale of the intrinsic dynamics of the electronic system. As can be seen from equations~\eqref{eq:EFG_dens_expand} and \eqref{eq:EFG_pheno}, a possible time-dependence of the strain $\varepsilon_{\alpha\beta}$ or the electric field $E_\gamma$ is not entering the derivation. Therefore, the obtained results are only valid in the adiabatic approximation for slowly varying strain or electric field, i.e. if the assumption, that the electronic system remains in its adiabatic ground state during evolution, is justified.

Given the functional dependence of the EFG tensor in a linearized adiabatic approximation on the electric field or strain, we immediately see that the time dependence of the EFG tensor is inherited from the time dependence of the external quantity. Thus, an application of an electric field $E(t) = E_0 \cos(\omega t)$ with a frequency in the radio-frequency regime, matching the transition frequency of nuclear spin states, may be used to locally modulate the EFG and thereby control nuclear spin state occupations coherently. This scheme has already been verified experimentally for a $^{123}$Sb (spin 7/2) nucleus in silicon~\cite{original_nature}. The interaction strength is determined by the intensity of the electric field and the quadrupole coupling tensor. Note that this scheme enables a pure electric control of the nuclear spin system without the need for oscillating magnetic fields in the radiofrequency regime.

\section{List of approximations in the model}\label{sec:A6}
For convenience, all approximations applied in the main part of this article are summarized and listed below:

\begin{itemize}
    \item Born approximation for total density matrix, i.e. decomposable density matrix throughout time-evolution
    \item negligible effect of spin system on electronic system due to relatively small coupling
    \item negligible effect of hyperfine coupling between electron spin and nuclear spin
    \item adiabatic approximation for the electronic system
\end{itemize}

All of these approximations are reasonable if the interaction of spin system and electronic system can be considered small and the time-dependence of the external field is quasi-static in comparison to typical time-scales of the electronic system. Both conditions are well satisfied in the case of the nuclear quadrupole interaction for NER and NAR. However, these conditions are also satisfied for excitations of the electronic system, as it is shown for the ONER protocol in Section~\ref{sec:oNER}.

\end{appendix}

\begin{acknowledgments}
Financial support by the Austrian Science Fund (FWF) under Grant P-36903N is gratefully acknowledged. We further thank the IT Services (ZID) of the Graz University of Technology for providing high performance computing resources and technical support.
\end{acknowledgments}

%\bibliography{main}
\input{main.bbl}

\end{document}